\documentclass[aps, preprintnumbers, tightenlines, nofootinbib, 11pt]{revtex4-2}

\usepackage{amssymb,amsmath,amsfonts,amsbsy,epsfig,wrapfig,caption,cancel,graphicx, pstricks, slashed, mathtools,booktabs,float}

\newcommand{\eq}[2]{\begin{equation}\label{#1}#2 \end{equation}}

\newcommand{\sm}[2]{\left|\begin{smallmatrix} {#1}\\ {#2}\end{smallmatrix}\right|}

\newcommand{\rd}[1]{{#1}}
\newcommand{\rdd}[1]{{\color{red}#1}}
\newcommand{\bl}[1]{{\color{blue}#1}}

\usepackage{xcolor}

\def\bea{\begin{eqnarray}}
\def\eea{\end{eqnarray}}
\def\be{\begin{equation}}
\def\ee{\end{equation}}
\def\ba{\begin{array}}
	\def\ea{\end{array}}

\def\nn{\nonumber}

\begin{document}
\author{Andrew D. Jackson$^{1}$}
\email{jackson@nbi.ku.dk}
\author{Subodh P. Patil$^{2}$}
\email{patil@lorentz.leidenuniv.nl}
\affiliation{1) Niels Bohr International Academy,\\ The Niels Bohr Institute, Blegdamsvej 17,\\ Copenhagen, DK 2100, Denmark\\}
\affiliation{2) Instituut-Lorentz for Theoretical Physics, Leiden University, 2333 CA Leiden, The Netherlands\\}
\date{\today}

\title{Phases of Small Worlds: A Mean Field Formulation}
\begin{abstract}
A network is said to have the properties of a small world if a suitably defined average distance between any two nodes is proportional to the logarithm of the number of nodes, $N$. In this paper, we present a novel derivation of the small-world property for Gilbert-Erd\"os-Renyi random networks. We employ a mean field approximation that permits the analytic derivation of the distribution of shortest paths that exhibits logarithmic scaling away from the phase transition, inferable via a suitably interpreted order parameter. We begin by framing the problem in generality with a formal generating functional for undirected weighted random graphs with arbitrary disorder, recovering the result that the free energy associated with an ensemble of Gilbert graphs corresponds to a system of non-interacting fermions identified with the edge states. We then present a mean field solution for this model and extend it to more general realizations of network randomness. For a two family class of stochastic block models that we refer to as dimorphic networks, which allow for links within the different families to be drawn from two independent discrete probability distributions, we find the mean field approximation maps onto a spin chain combinatorial problem and again yields useful approximate analytic expressions for mean path lengths. Dimorophic networks exhibit a richer phase structure, where distinct small world regimes separate in analogy to the spinodal decomposition of a fluid. We find that is it possible to induce small world behavior in sub-networks that by themselves would not be in the small-world regime.
\end{abstract}
\maketitle
\tableofcontents

\section{Preliminaries}
Although small-worlds networks have proved to be fertile ground for theoretical musing among mathematicians and scientists, the initial recognition and exploitation of the small-worlds phenomenon has far more empirical origins. The existence of such networks was first suggested by the Hungarian journalist and author Frigyes Karinthy in a short story published in 1929\footnote{The story, ``Chain-Link'', appeared in a collection entitled {\em Everything is Different\/}.  An English translation of this story is available online at \url{http://vadeker.net/articles/Karinthy-Chain-Links_1929.pdf}}.  In this story Karinthy asserts that, ``using no more than five individuals'', an unbroken link of acquaintances can be established to connect any two people in the world.  This is the source of the now-familiar notion of ``six degrees of separation''.  Although Karinthy offered anecdotal evidence for his assertion, forty years passed before Travers and Milgram provided empirical evidence for this assertion and coined the term ``small worlds'' \cite{travers1969s}. The small-worlds property has since become a widely-used empirical tool for exploring the relational structures of a variety of social, biological, and physical systems.  Such empirical studies have benefited greatly from the availability of massive data sets, among the largest of which is a snapshot of the approximately 700 million monthly active users of Facebook in 2011 (a figure that has since ballooned to 2.7 billion users in 2020) \cite{2011arXiv1111.4503U}. Not only does this social network exhibit the properties of a small world, it turns out that Facebook users on average have only four degrees of separation between one another \cite{2011arXiv1111.4570B}. Small world networks, and their associated  properties manifest in a multitude of other contexts as well, with recent experience proving its relevance to epidemiology \cite{newman2002spread, keeling2005networks, miller2009percolation}. Similarly, many professional networks from Broadway \cite{uzzi2005collaboration} \rd{ and Wall Street \cite{doi:10.1177/14761270030013002} to scientific collaborations \cite{PhysRevE.64.026118}} exhibit small world properties, encapsulated for mathematicians and theoretical physicists by their Erd\"os number, for example.

The first mathematical contribution to the small-worlds problem was provided in 1959 in the form of Erd\"os-Renyi random graphs \cite{erdos59a}, a class that was proposed independently by Gilbert \cite{gilbert1959}. Such graphs correspond to a set on $N$ nodes where a link between any two chosen nodes will exist with probability $p$.  Erd\"os-Renyi-Gilbert random graphs represent a limiting case of randomness in contrast to regular lattices, where links to nearest neighbors are to some extent guaranteed.  On a regular lattice-like network, the average distance between any two nodes scales with some power the number of nodes, $N$, in the network.  Along with other small-worlds networks\footnote{\rd{Scale-free networks, i.e. networks whose degree distributions exhibit an asymptotic power law behavior $p(k) \propto k^{-\gamma}$ with $k$ denoting the degree (see e.g. \cite{doi:10.1126/science.286.5439.509}) have furthermore been shown to be \textit{ultra-small}, where average distances scale as $\log(\log N)$ provided $\gamma < 3$ \cite{cohen2003scale}.}}, Erd\"os-Renyi-Gilbert random graphs have the property that the distance between nodes is proportional to $\log N$. Interpolating between these two extremes is a class of random networks introduced by Watts and Strogatz \cite{1998Natur.393..440W}, for which a regular ring lattice with links to $k$ nearest neighbors can have any given link reassigned to a random node with probability $p$. A variant of such networks was studied by Newman and Watts \cite{NW}. Instead of reassigning links, random links can be inserted between pairs of nodes. Both classes of models \cite{1998Natur.393..440W, NW} exhibit small-world scaling above a critical probability $p\, \rd{\sim}\, 1/N$.  The regularity inherited from the original lattice before link reconnections or insertions are performed makes it possible to demonstrate small-world scaling analytically in the large $N$ limit. 

There is a qualitative distinction however between the small-world scaling seen in Erd\"os-Renyi-Gilbert and Watts-Strogatz classes of networks.  The former exhibit a smaller clustering coefficient than the latter, whereas many real world examples of small-world networks exhibit large clustering coefficients \cite{1998Natur.393..440W}. \rd{Realizations of random networks that model a wider range of real systems can be obtained by specifying different degree distributions, by which one can analytically study the distribution of shortest path lengths \cite{Dorogovtsev_2003, 2004math......7092V, van2008universality, nitzan2016distance, melnik2016simple}, as well as the shell (i.e. differential) \cite{shao2008fractal, shao2009structure, katzav2015analytical} and cycle structure \cite{bonneau2017distribution}. In many contexts of interest, networks are not static entities but rather grow over time. Analytic studies of the distribution of shortest paths have also been made in the context of network growth via node duplication, as detailed in \cite{steinbock2017distribution, steinbock2019analytical}.}

Erd\"os-Renyi graphs are nevertheless useful in understanding the properties of real world networks through the variety of analytical results that can be proved for them\footnote{\rd{Remarkably, the simple abstraction represented by an Erd\"os-Renyi network may even offer a model for certain macroscopic neuronal oscillations \cite{10.3389/fncir.2015.00065}.}}. Perhaps the most celebrated of these is the critical probability found in the semi-regular network structure of \cite{NW} also exists in the purely random context \cite{erdos59a, gilbert1959}.  There exists a threshold value of $p_c \rd{\sim} 1/N$ above which statements about any representative graph are almost always true.  It can be demonstrated, for example, that for \rd{$p > (\log N)/N$} there will be a `giant component' containing all nodes in the large $N$ limit\cite{erdos59a,Erdos:1960,ER}\footnote{\rd{As we will see below, when $p = \kappa/N$ the number of particles \textit{not} in the giant component is $\mathcal O(N)$. When $\kappa \to \log N$, the number of nodes not in the giant component is $\mathcal O(N^0)$. In other words, past the critical edge probability $p_* = (\log N)/N$, there are no longer any finite clusters in the large $N$ limit.}}. The emergence of such a hub structure is central to the small-world scaling of large networks\footnote{\rd{However, many interesting properties can also be inferred for sub-critical Erd\"os-Renyi networks, such as the degree distribution of the giant component \cite{tishby2018revealing}, and the distribution of shortest paths across all components \cite{katzav2018distribution}.}}. However, although one can demonstrate small-world scaling of the average distance between nodes, $\ell$, in an Erd\"os-Renyi-Gilbert network in limiting cases where \rd{$p \gg (\log N)/N$} (or vice versa), an analytic understanding of the approach to the transition point for $\ell$ remains elusive.

Furthermore, although the semi-regularity of the Watts-Strogatz \rd{class of} networks permits a continuum generalization for which a variant of the mean field approximation has been proposed \cite{Newman_2000}, direct application of such methods fails for Erd\"os-Renyi-Gilbert graphs because of the lack of an underlying lattice sub-structure\footnote{See however \cite{PhysRevE.90.062801} for a network model with link randomness that interpolates between the Watts-Strogatz and Erd\"os-Renyi-Gilbert models.}. In this paper, we present another variant of the mean field approximation, not predicated on a continuum generalization of the underlying graph structure that allows for an analytic demonstration of small world scaling in Erd\"os-Renyi-Gilbert random networks. In the case of Watts-Strogatz graphs \cite{1998Natur.393..440W,NW}, the underlying lattice structure results in a phase transition at $p = 1/N$ from an $\ell \propto N$ scaling to an $\ell \propto \log N$ scaling. For Erd\"os-Renyi-Gilbert graphs, the \rd{(percolation)} transition occurs at a critical \rd{$p_c = 1/N$ \cite{https://doi.org/10.1002/rsa.3240060204, molloy1998size, newman2018networks}} and is of a somewhat different nature due to the absence of an underlying lattice. Here, the transition is from a regime characterized by the existence of multiple disconnected components of similar size (naturally identifiable as `domains') to a regime dominated by a single \rd{giant connected component}. Moreover, we can show that the mean field approximation fails close to the critical point between these two regimes even in the large $N$ limit due to its inability to cope with the inherent inhomogeneities when the network exists in a phase composed of multiple disconnected components. However, the utility of the approximation described here, characteristic of any mean field approximation, is that it offers analytic understanding of a restricted class of phenomena.  Fortunately, the emergence of small-world scaling away from the critical point is among them. 

The assumption that all nodes in a network have the same probability of forming links is evidently an idealization of real networks in which some nodes can have an increased probability of forming links.  This is the case, e.g., when considering co-author linkages within a given academic field.   Paul Erd\"os had some 240 co-authors, which is unusually large in mathematics.  The ability of such special nodes to reduce the path lengths was noted by Karinthy in 1929:  ``It's always
easier to find someone who knows a famous or popular figure than some insignificant person''.  Fortunately, the mean field approximation is readily generalized to describe graphs containing such inequivalent nodes, and it will be shown that qualitatively new network behavior emerges from this extension. We consider a simple two family stochastic block variant of the Erd\"os-Renyi-Gilbert network, where links within a given family appear with fixed but distinct probabilities and links between the families appear with some finite probability and allowing for the relative sizes of these families to be arbitrary. This is not quite the widely studied \textit{planted partition} stochastic block model, as we allow for the links within members of each family to appear with different probabilities, and so we will refer to this variant in what follows as a \textit{dimorphic network}. 

The dimorphic variant of the Erd\"os-Renyi-Gilbert network exhibits a richer phase structure than the standard case and is also accurately modelled by the mean field approximation provided that the network does not consist of ``too many'' disconnected components. One phase corresponds to the situation where the network factorizes into two strongly connected sub-networks with a limited number of `bridge' links between them. The transition to this phase from a more homogeneous, but still disordered phase is roughly analogous to the spinodal decomposition of fluids. We also find that it is possible to induce small world behavior in a sub-network even if it does not meet the criteria to be in the small-world regime by itself, provided it connects to a sub-network that does. One can also envision a variant of the dimorphic network where the links within and between sub-networks have different weights.  The formalism described here can readily be extended to describe such, and we shall pursue this possibility in subsequent investigations.

The remainder of this paper is organized as follows.  In section II, we construct a formal generating functional for an arbitrary undirected weighted network, recovering the result that the associated free energy for an ensemble of Gilbert graphs corresponds to a system of non-interacting fermions residing at the edges. We also show how one can obtain formal expressions for ensemble averages of various network observables such as the average degree of the network, the degree distribution as well as the distance matrix.  We note, however, that actually evaluating these quantities will be unduly cumbersome in practice.  In section III we introduce a mean field approximation that facilitates the evaluation of the ensemble average of the distance matrix for Erd\"os-Renyi-Gilbert networks and demonstrate that it accurately reproduces the small worlds scaling away from the critical edge probability \rd{$p_* = (\log N)/N$}. We show that this approximation fails as one approaches the latter critical point, and that this is to be expected since the network fragments into multiple domains. In sections IV and V, we generalize the mean field approximation to dimorphic networks and consider its different phases.  The various technical details will be relegated to appendices.  We offer concluding remarks in section VI, where we discuss generalizations of the mean field approximation presented here to more general varieties of network disorder.

\section{Random networks from random matrices}
Various properties of a network can be obtained from its associated adjacency matrix, \textbf{A}.  All diagonal elements are 0. For unweighted networks, the off-diagonal element $\textbf{A}_{ij}$ is 1 if nodes $i$ and $ j$ are connected by an edge and otherwise 0.  For weighted networks the non-vanishing entries of $\textbf{A}_{ij}$ can be any real positive number\footnote{Here, we will not consider networks where nodes can have edges that begin and end on the same node (i.e. loops) and will thus have adjacency matrices with diagonal entries.}. Undirected networks have symmetric adjacency matrices; directed networks do not. It is clear that random networks can be described by the associated ensemble of random adjacency matrices defined by a particular probability measure.

Given a probability measure $P(\textbf{A})$ over the space of $N\times N$ real matrices    
\eq{pm}{\int \mathcal D \textbf{A} P[\textbf{A}] = 1}
where 
\eq{meas}{\mathcal D\textbf{A}P[\textbf{A}] := \prod_{i,j} d A_{ij} P_{ij},}
we can define the generating functional
\eq{}{Z[J] = \int \mathcal D\textbf{A}P[\textbf{A}] e^{-{\rm Tr\,}A J^T} }
such that expectation values of arbitrary powers of the adjacency matrix can be obtained from successive derivatives of $Z[J]$
\eq{moments}{\langle \textbf{A}^m_{ij} \rangle = (-1)^m\frac{\delta^m Z[J]}{\delta J_{ii'}\delta J_{i'k}... \delta J_{l l'}\delta J_{l'j} }\Bigg|_{J\equiv 0}.}
The Gilbert-Erd\"os-Renyi model \cite{gilbert1959} corresponds to an unweighted, undirected network of $N$ nodes, where a link between any two nodes occurs with a fixed probability $p$. The associated probability measure is given by\footnote{We also illustrate a variant of the Watts-Strogatz model \cite{1998Natur.393..440W} in Appendix A.}
\begin{eqnarray}
\label{G}
(i < j)~~~~~~ P_{ij} &=& p\,\delta(A_{ij} - 1) + (1-p)\,\delta(A_{ij})\\ \nn
(i = j)~~~~~~ P_{ij} &=& \delta(A_{ii})\\ \nn 
(i > j)~~~~~~ P_{ij} &=& \delta(A_{ij} - A_{ji})
\end{eqnarray}
Computation of the generating functional $Z[J]$ for any given probability measure is straightforward since the measure and the source term factorize as
\eq{zans}{Z[J] =  \prod_{i,j}\int\,d A_{ij}\, P_{ij}e^{- A_{ij}J_{ij}}\ .}
It follows directly from Eq. \ref{G} that
\eq{GZ}{Z[J] = (1-p)^{\frac{N(N-1)}{2}}\prod_{i < j}\left[1 + e^{-\left(\lambda + J_{(ij)}\right)} \right];~~~\lambda := - \log\left(\frac{p}{1-p}\right),~J_{(ij)} := J_{ij} + J_{ji}\ .}
We can also define the generating function, $W[J]$, for all connected correlation functions as
\eq{}{e^{-W[J]} := Z[J]} 
which from Eq. \ref{GZ} is given by
\eq{GW}{W[J] = \frac{N(N-1)}{2}\log\left(\frac{1}{1-p}\right) - \sum_{i < j} \log\left[1 + e^{-(\lambda + J_{(ij)})} \right].}
Since the pre-factor in Eq. \ref{GZ} serves only to normalize $Z[0] = 1$, it can be discarded provided that all moments are obtained by dividing Eq. \ref{moments} by $Z[0]$. The generating function for Gilbert random graphs then becomes 
\eq{GZ0}{Z[J] = \prod_{i < j}\left[1 + e^{-\left(\lambda + J_{(ij)}\right)} \right]\ {\rm with}\ \lambda := - \log\left(\frac{p}{1-p}\right)\ \ {\rm and}\ \ J_{(ij)} := J_{ij} + J_{ji}}
with a corresponding free energy $W[J]$ given by
\eq{GW}{W[J] = - \sum_{i < j} \log\left[1 + e^{-(\lambda + J_{(ij)})} \right]\ .}
We note that the above corresponds to the free energy for an `exponential random graph' obtained from a probability measure that maximizes the Gibbs entropy subject to a particular set of constraints \cite{PN}. As noted in \cite{PN}, the partition function Eq. \ref{GZ0} and free energies Eq. \ref{GW} correspond to a system of non-interacting fermions localized on the links. This is made even more apparent when one calculates the one particle occupation numbers:
\eq{}{-\frac{1}{Z}\frac{\delta Z}{\delta J_{ij}}\Bigg|_{J\equiv 0} \hspace{-10pt}= n_{ij} = \frac{1}{e^\lambda + 1} = p\ .}
Taking the appropriate derivatives of Eq. \ref{GZ0} with respect to the source components $J_{ij}$, one can obtain ensemble averages of any network observable that can be expressed in terms of the adjacency matrix. In the case of unweighted networks for instance, the average number of paths of length $k$ connecting nodes $i$ and $j$, defined as $\#_{ij}(k)$ is given precisely by Eq. \ref{moments}:
\eq{}{\#_{ij}(k) = \langle \textbf{A}^k_{ij} \rangle.}
The degree of the i$^{th}$ node in any given network realization is given by the bilinear form $(e_i,\textbf{A} u)$ where $e_i$ is the i$^{th}$ unit vector, and where $u = \sum e_j$ is the vector with 1 in all of its entries. Therefore the ensemble averaged degree of the i$^{th}$ node, defined as $k_i$, is given by
\eq{}{k_i = \langle(e_i,\textbf{A} u)\rangle = \sum_{j=1}^N \langle \textbf{A}_{ij} \rangle\ .} 
Note that this need not be the same for all $i$ as the $P_{ij}$ in Eq. \ref{meas} need not be the same for all links connecting nodes $i,j$. The average degree of the entire network, $\langle k \rangle$, can be obtained as
\eq{}{\langle k \rangle = \frac{1}{N}\langle(u,\textbf{A} u)\rangle = \frac{1}{N}\sum_{i,j=1}^N \langle \textbf{A}_{ij} \rangle\ .} 
The analogs of the above expressions for general heterogeneous networks can also be found, for instance in \cite{Squartini_2015}\cite{cimini2019statistical}.

The average value of the shortest path length between any two nodes is of particular interest in this investigation. This first requires us to specify the notion of a distance on a network. For a general weighted network, a natural (although by no means unique) choice is provided by the analog of effective conductance in an electrical circuit \cite{OPSAHL2010245}:
\eq{metric}{D_{ij} = \min\left\{ \left(\frac{1}{A_{ih}} + ... + \frac{1}{A_{hj}}  \right)\right\}}
where the sum includes all possible paths connecting nodes $i$ and $j$. For the unweighted networks to be considered here, this reduces to the usual notion of the minimum number of links connecting nodes $i$ and $j$. In this case, the diagonal elements of the distance matrix \textbf{D} are defined as 0, and the off-diagonal elements are smallest number of links connecting any pair of nodes (which can be infinite if no such path exists).  The average distance between any two nodes is thus given by
\eq{elld}{\ell = { N\choose 2}^{-1} \sum_{i<j} \langle \textbf{D}_{ij}\rangle  = { N\choose 2}^{-1}\langle (u,\textbf{D} u)\rangle\ .}
Proving the small worlds property for a given class of random networks thus consists in proving the limiting behavior
\eq{}{\lim_{N\to \infty} \ell = \lim_{N\to \infty} { N\choose 2}^{-1}\sum_{i<j} \langle \textbf{D}_{ij}\rangle \to \log N\ .} 
Depending on the particular realization of network randomness, the resulting distance matrix can prove to be disconnected.  This means that $\textbf{D}$ can be brought into block diagonal form by a similarity transform involving a permutation matrix  The individual blocks will correspond to connected component graphs, and the off-diagonal block entries will be infinite. In such cases, it seems natural to define an average distance exclusively on the basis of sums performed within the distinct connected components. 

There are a variety of algorithms for obtaining the distance matrix of a particular network given its adjacency matrix, for instance Dijsktra's algorithm and its variants \cite{EWD}. In the special case of an unweighted network, however, one can also obtain a formal expression for the distance matrix in terms of the adjacency matrix. Given that the $k^{th}$ power of the adjacency matrix has entries $\textbf{A}^k_{ij}$ that count the number of paths of length $k$ between nodes $i$ and $j$, the shortest distance between any two nodes $i, j$ is given by the first power $k$ for which $\textbf{A}^k_{ij} \neq 0$. Therefore, 
\eq{Ddef}{\textbf{D}_{ij} = \min \{k\,|\, \textbf{A}^k_{ij} > 0\}  }
An analytic approximation to the minimum function (exact in the limit $\Lambda \to \infty$) is given by
\eq{}{\textbf{D}_{ij} = \lim_{\Lambda \to \infty} -\frac{1}{\Lambda} \log\sum_{q=1}^\infty \Theta(\textbf{A}^q_{ij} - 1/2)\,e^{-\Lambda q} }
where the offset of the Heaviside step function could be any real number between 0 and 1. In particular, we wish to calculate the ensemble expectation value of the above expression, that is 
\eq{minD}{\langle\textbf{D}_{ij}\rangle = \lim_{\Lambda \to \infty} -\frac{1}{\Lambda}  \langle \log\sum_{q=1}^\infty \Theta(\textbf{A}^q_{ij} - 1/2)\, e^{-\Lambda q} \rangle\ .}

As it stands, Eq. \ref{minD} is calculationally cumbersome. It is tempting to bring the expectation value in Eq. \ref{minD} inside the logarithm by perhaps assuming that network disorder is \rd{``annealed''}, or (by some abuse of terminology) invoking a variant of the replica trick \cite{1987sgtb.book, Edwards_1975}. It is unclear if either assumption is justified. Moreover, with the exception of Edwards-Anderson type (Gaussian) disorder for weighted networks \cite{Edwards_1975} (requiring the more general metric Eq. \ref{metric}), it is unclear if any calculational advantage is to be gained. Rather, in the next section we will suggest a way to approximate the above for the special case of unweighted networks using a particular variant of the mean field approximation.   

\section{Small worlds --- a mean field description}

In this section we consider a ``mean field'' description of Gilbert-Erd\"os-Renyi random networks.  Although this description is not exact, it will be seen to provide a semi-quantitative description of their properties.  For the case of unweighted networks, we see from Eq. \ref{Ddef} that the elements of the distance matrix $\textbf{D}_{ij}$ give the length of the shortest path between nodes $i$ and $ j$.  This is simply the smallest integer $k$ such that $\textbf{A}^k_{ij}$ is non-zero. It should be clear that no such shortest path can involve any repeated indices in the product of the $k$ adjacency matrices since this would correspond to backtracking over a previous step. The determination of the distribution of shortest path lengths involves two steps.  First,  we must determine the probability that $\textbf{A}^k_{ij}$ is non-zero for a given value of $k$.  This is given as one minus the probability that $\textbf{A}^k_{ij}$ vanishes subject to the constraint that there are no repeated indices in the intermediate summations within the matrix product.  One simply computes the probability that the string $A_{i i_1}A_{i_1 i_2}... A_{i_{k-1}i_k}$ vanishes, where $i_1 ... i_{k-1}$ are summed over subject to the constraint that no index is repeated. From this, one sees immediately that the probability, $p_k$, that there will be a shortest, non-backtracking path\footnote{In what follows, the adjective \textit{shortest} should be taken to refer to shortest, non-backtracking paths. The subsequent expressions rely on the approximation that different shortest paths of length $k$ represent independent events --- as discussed further (cf. footnote \ref{MFf}), this relates to the assumptions that underlie our mean field approximation.} of length $k$ between the nodes $i, j$ is given by 
\eq{pk}{p_k = 1 - (1 - p^k)^{\gamma_k}}
where 
\eq{gdef}{\gamma_k = (N-2)!/(N-k-1)! \ .}
The probabilities $p_k$ represent a true ensemble average for the unweighted network
.  The second part of the calculation requires us to modify the $p_k$ by imposing the constraint that there is no path of length less than $k$ that connects the nodes in question.  Since it is difficult to impose this constraint exactly, we will proceed as follows.  The ensemble average probability, $q_k$, that there does not exist a shortest path of length $k$ is  $1 - p_k$.  Using this ensemble averaged quantity, the probability that the shortest path will be exactly of length $k$ (and no shorter), denoted as $P(k)$ can be estimated as 
\eq{Pk}{P(1) = p, {~~\rm and~~} P(k) = p_k \prod_{n=1}^{k-1}q_n,~~ 1 < k < N.}

We note that there is also some probability, $P_{\infty}$, that there will be no path of any length connecting two nodes.  This probability is readily seen to be 
\eq{Pinf}{P_{\infty} = \prod_{n=1}^{N-1}\, q_n.}  
\rd{As we shall see shortly, this quantity is directly relates to the size of the largest connected component (c.f. the quantity $F_\infty$ in \cite{katzav2015analytical}).} With the inclusion of this additional term (which can be $0$), the probability distribution function (PDF) will have the correct normalization
\eq{norm}{\sum_{k=1}^{N-1} P(k) + P_\infty = 1.}  
These results would be exact if the matrices $\textbf{A}^k$ for various values of $k$ were uncorrelated.  Unfortunately, this is not the case.  The operation of ensemble averaging does not commute with other aspects of the computation and should be performed last.  It is thus an approximation to make use of ensemble averaged values of $q_k$ in the calculation of the $P(k)$, and it remains to be seen whether this approximation is reliable\footnote{\label{MFf} It is in this sense that the treatment in this section amounts to a mean field approximation.}. We note, however, the strong similarity of this approximation to the mean field approximation familiar from statistical mechanics and many-body physics.  Mean field results can be useful provided that the system is homogeneous with correlations/fluctuations that are sufficiently small. Determination of the degree of reliability of mean field results in the present case will require comparison with simulations.  

In the following examples of mean field results and their comparison with the results of simulations, we will consider the choice 
\eq{kappa}{p = \frac{\kappa}{N-1}.}
This is motivated by our wish to model systems for which the properties of the individual nodes (e.g., the average number of links per node, $\kappa$) are independent of the size of the system.  Although more elaborate choices of $p$ might be appropriate for specific systems, their adoption would require physical justifications that are beyond the scope of our investigation. The simulations were done over ensembles consisting of $10^2$ samples, and converge rapidly away from the critical point. 

\begin{figure}[t!]
		\begin{center}
			\includegraphics[height=3in, width = 4.2in]{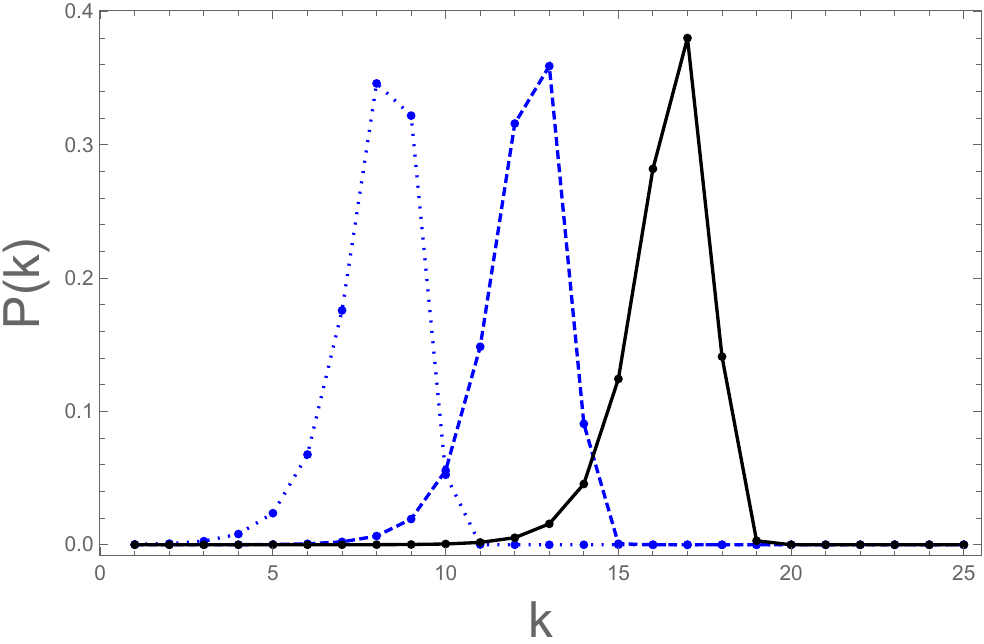}
			\captionof{figure}{Mean field values of  $P(k)$ with $p = 3/(N-1)$ for $N = 10^4$ (blue dotted), $10^6$ (blue dashed), and $10^8$ (black).  Using Eq.\,\ref{km} we find that $x_{\rm mode}  = 7.93, 12.12$ and $16.32$ respectively, consistent with $k_{\rm mode}$ being identified by $[x_{\rm mode}]_\pm$ as discussed below Eq. \ref{km}. \label{mono_plots}}
		\end{center}
\end{figure}

Fig.\ \ref{mono_plots} shows the values of $P(k)$ determined by Eqs.\ \ref{pk} - \ref{Pk} for several values of $N$.  Consistent with small-worlds expectations, the peak of the distribution increases logarithmically with $N$ and its width is roughly independent of $N$.  Thus, relevant features of this distribution can be determined from a number of terms proportional to  $\log N$. Although we have verified the validity of the mean field approximation for different values of $N$ up to $ N = 10^3$, note that it very challenging to perform simulations for values of $N$ as large as those shown in Fig.\ \ref{mono_plots}.  This figure provides a strong indication of small worlds behavior for this choice of $\kappa$, although we shall see below that this result is not true in general.  

An analytic expression of how the mean field description captures these essential features of the distribution of path lengths can be obtained by considering the maximum value (i.e., the mode) of the $P(k)$ distribution for large $N$.  In fact, this expression will also tell us when we are in the small-worlds domain.  From Eq. \ref{Pk} we see that 
\eq{cond}{\frac{P(k+1)}{P(k)} = p_{k+1}\left(\frac{1}{p_k} - 1\right).}
The distribution will have an extremum whenever this ratio is equal to one.  Imposing this condition  implies that 
\eq{max}{1 = \left[2 - (1 - p p^k)^{\gamma_k (N-k-1)}\right](1 - p^k)^{\gamma_k}\,.}
Replacing $k$ with the continuous variable $x$, using the analytic approximation
$\Gamma(N + \alpha)/\Gamma(N) \to N^\alpha$ as $N\to\infty$, and using the fact that, for example, 
\begin{displaymath}
	(1 - p^{x+1})^{\gamma_{x+1}} \approx e^{-\gamma_{x+1}p^{x+1}}\ \ {\rm for}\ \ p < 1\ ,
\end{displaymath}
we see that Eq. \ref{max} is equivalent to the condition
\eq{tayl}{1 = \frac{1}{2}\left(e^{\frac{\tau}{\kappa}} + e^{-\tau}  \right) \ \ {\rm with}\ \ \tau =  p\kappa^x = \frac{\kappa^{x+1}}{N-1}\ .}
Numerically evaluating the root of the above for $\tau$ as a function of $\kappa$, and solving for $x$ which we identify as $x_{\rm mode}$ results in\footnote{We note the formal similarity of Eq. \ref{km} with Eqs. 1 and 3 in \cite{PhysRevE.72.026108} which consider pairwise mean path lengths conditional on the pair of nodes in question having a particular degree, the function $r(\kappa)$ evidently quantifying the dependence under the identification $\kappa = \langle k \rangle$.}
\eq{km}{x_{\rm mode}(\kappa,N) = \frac{1}{\log \kappa} \log (N-1) + \frac{1}{\log \kappa}\log r(\kappa) - 1,~~~~ k_{\rm mode} = [x_{\rm mode}]_\pm}  
where $r(\kappa)$ is the root of Eq. \ref{tayl}. Given that $k$ is a discrete variable, whose continuous generalization $x$ attains a maximum at some value, the most we can conclude is that $k_{\rm mode}$ is either of the nearest integers rounded up or down around the value obtained from Eq. \ref{km}, denoted $[x_{\rm mode}]_\pm$. As seen in Fig. \ref{mono_plots}, this provides a reasonably accurate determination of the  mode of the PDF given by Eq. \ref{Pk}.  Note, however, the singular behaviour of $k_{\rm mode}$ when the average number of links per node is $\kappa \le 1$.  This is related to the fact that, for sufficiently small $\kappa$, the $P(k)$ decrease monotonically with $k$ and no local maximum exists, due to the fragmentation of the network into disconnected components. \rd{Moreover, we note that Eq. \ref{km} diverges as $1/(\kappa - 1)$ as $\kappa \to 1$ from above, in agreement with the divergence found for the mean path lengths in \cite{katzav2018distribution}.}

\begin{figure}
	\begin{center}
		\includegraphics[height=2.9in,width = 4.0in]{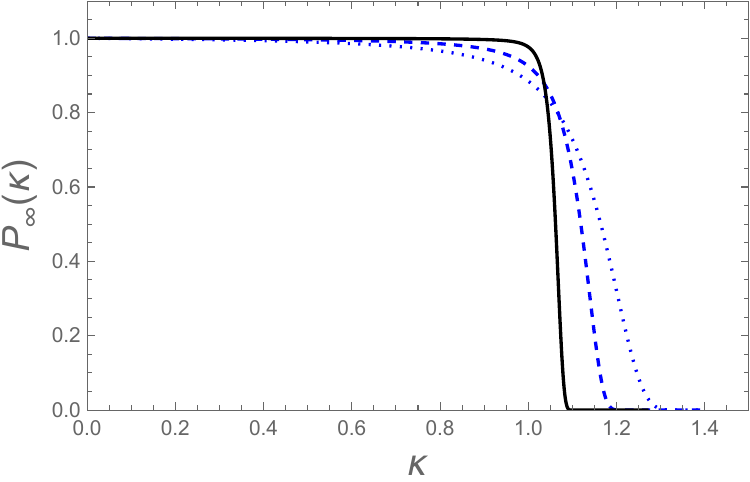}
		\caption{$P_{\infty}$ as a function of $\kappa$ for $N =$ 100 (dotted), 250 (dashed) and 2500 (solid).\label{order}}
	\end{center}
\end{figure}

To see this more clearly, Fig.\ \ref{order} shows the dependence of $P_{\infty}$ on the average number of links per node, $\kappa$, as determined by the mean field result Eq.\ \ref{Pinf}.  As one passes through the critical edge probability \rd{$p_* = (\log N)/N$ (which we rederive shortly)}, mean field results suggest that the composition of the network undergoes a significant change \rd{-- an expected result from the perspective of percolation theory \cite{https://doi.org/10.1002/rsa.3240060204, molloy1998size, newman2018networks}}. For $p > p_*$, essentially all nodes are linked to one another in a genuine small world state.  For $p < p_*$ the network exists in a phase composed of a multitude of clusters of small size.  In such a phase, there will be a significant number of nodes that are not linked.  In other words, the mean field approximation predicts that $P_{\infty} \to 1$ for $\kappa \ll 1$ and $P_{\infty} \to 0$ for $\kappa \gg 1$. The mean field values of $P_{\infty}$ are shown in Fig. \ref{order} as a function of $\kappa$ for several values of $N$.  We see from the figure that this transition becomes increasingly sharp with increasing $N$. It is tempting to regard $P_\infty$, \rd{or rather $(1 - P_\infty)^{1/2}$} as an order parameter and to view this as a mean field indication of a phase transition at $\kappa = 1$.  However, mean field descriptions of phase transitions are known to be incorrect for many systems (e.g., the Ising model in one and two dimensions) as a consequence of its failure to account for significant fluctuations near a critical point, rendering meaningless any simple averaged descriptions of the system. In fact, we are obliged to consider an entirely different quantity not obtainable from our mean field approximation \rd{(although well understood in the context of percolation theory \cite{molloy1998size, newman2018networks})} as the order parameter pertaining to the phase transition in Erd\"os-Renyi-Gilbert random networks. We identify this order parameter in the next subsection, and turn to the question of how to interpret the mean field approximation in addition to comparing with simulations.

\subsection{Mean fields and simulations}
Before comparing the mean field results with simulations, we first consider the probability that a given node is a ``hermit'' with no links to any of the remaining $(N-1)$ nodes, which is one of the few results that can be determined exactly without the necessity of simulations. This probability is given by $(1 - p)^{N-1}$, or
\eq{hermit}{\left( 1 - \frac{\kappa}{N-1} \right)^{N-1},}
which approaches $\exp[-\kappa]$ in the limit of large $N$. This is an exact statement valid for all $\kappa$ and $N$.  Evidently, the number of nodes which have no connection with the largest small-worlds cluster must be greater than or equal to the number of such hermits. The latter can be similarly estimated via the assumption that for large enough $N$, the fraction of nodes contained in the largest cluster does not depend on $N$ (although it does depend on $\kappa$). Denoting this fraction as $f_{\rm cluster}$, we realize that if we  were to add one more node and if $f_{\rm cluster}$ is to remain unchanged, the probability that this new node not belong to the largest cluster must be $1 - f_{\rm cluster}$. That is
\eq{}{\left(1-\frac{\kappa}{N-1}\right)^{N f_{\text {cluster }}}=1-f_{\text {cluster }},}
which in the large $N$ limit implies
\eq{cop}{f_{\text {cluster }}=1-\exp \left[-\kappa f_{\text {cluster }}\right].} 
This is in fact the order parameter relevant to the percolation phase transition in Erd\"os-Renyi-Gilbert random networks \rd{\cite{https://doi.org/10.1002/rsa.3240060204, molloy1998size, newman2018networks}}. Unfortunately, this is not obtainable from our mean field approximation, which focuses on the links of the network as opposed to the number of nodes in the largest cluster. We note that Eq. \ref{cop} has only the trivial solution $f_{\rm cluster} = 0$ for $\kappa < 1$, whereas it also admits a non-trivial positive solution for $\kappa \geq 1$, which also depicts simulations of the largest cluster size. \rd{Eq. \ref{cop} has the formal solution \cite{tishby2018revealing}
\eq{}{f_{\rm cluster} = 1 + \frac{1}{\kappa}W\left(-\kappa e^{-\kappa}\right),}
where $W(z)$ is the Lambert $W$ function\footnote{Defined as the root of the equation $w e^w - z = 0$, i.e. $w = W(z)$}. However, by considering the quantity $\bar f_{\rm cluster} = 1 - f_{\rm cluster}$ and defining $\kappa = \log N$, one can rewrite Eq. \ref{cop} as 
\eq{fbeq}{{\bar f_{\rm cluster}}=\frac{1}{N} N^{{\bar f_{\rm cluster}}}.}
Starting with $\bar f_{\rm cluster} = 0$ and iterating the equation repeatedly, one obtains a power series expansion in $(\log N)/N$ and read off the general series coefficients to obtain the exact solution
\eq{}{ \bar{f}_{\rm cluster } = \frac{1}{N}+\frac{1}{N} \sum_{k=1}^{\infty} \frac{(k+1)^{k-1}}{k !}\left(\frac{\log N}{N}\right)^{k}, }  
from which it is clear that $\bar f_{\rm cluster} \to 0$, or $f_{\rm cluster} \to 1$ as $N \to \infty$.}

In order to compare the percolation order parameter $f_{\rm cluster}$ with the mean field order parameter $P_\infty$, we realize that in the large $N$ limit, the fraction of connected pairs of nodes is approximated to leading order by $f^2_{\rm cluster}$. Hence, the fraction that are not connected is given by $1 - f^2_{\rm cluster}$. Thus, by comparing $(1 - P_\infty)^{1/2}$ to $f_{\rm cluster}$, one can make a fairer determination as to whether the mean field order parameter accurately captures the nature of the phase transition\footnote{\rd{The main shortcoming of our mean field approximation is that it fails to condition on the giant component. When one does so, it has been shown exactly that $(1 - P_\infty)^{1/2} = f_{\rm cluster}$ (cf. Eq. 119 in \cite{tishby2018revealing}).}}.    
\begin{figure}
	\begin{center}
		\includegraphics[height=2.50in,width = 4.5in]{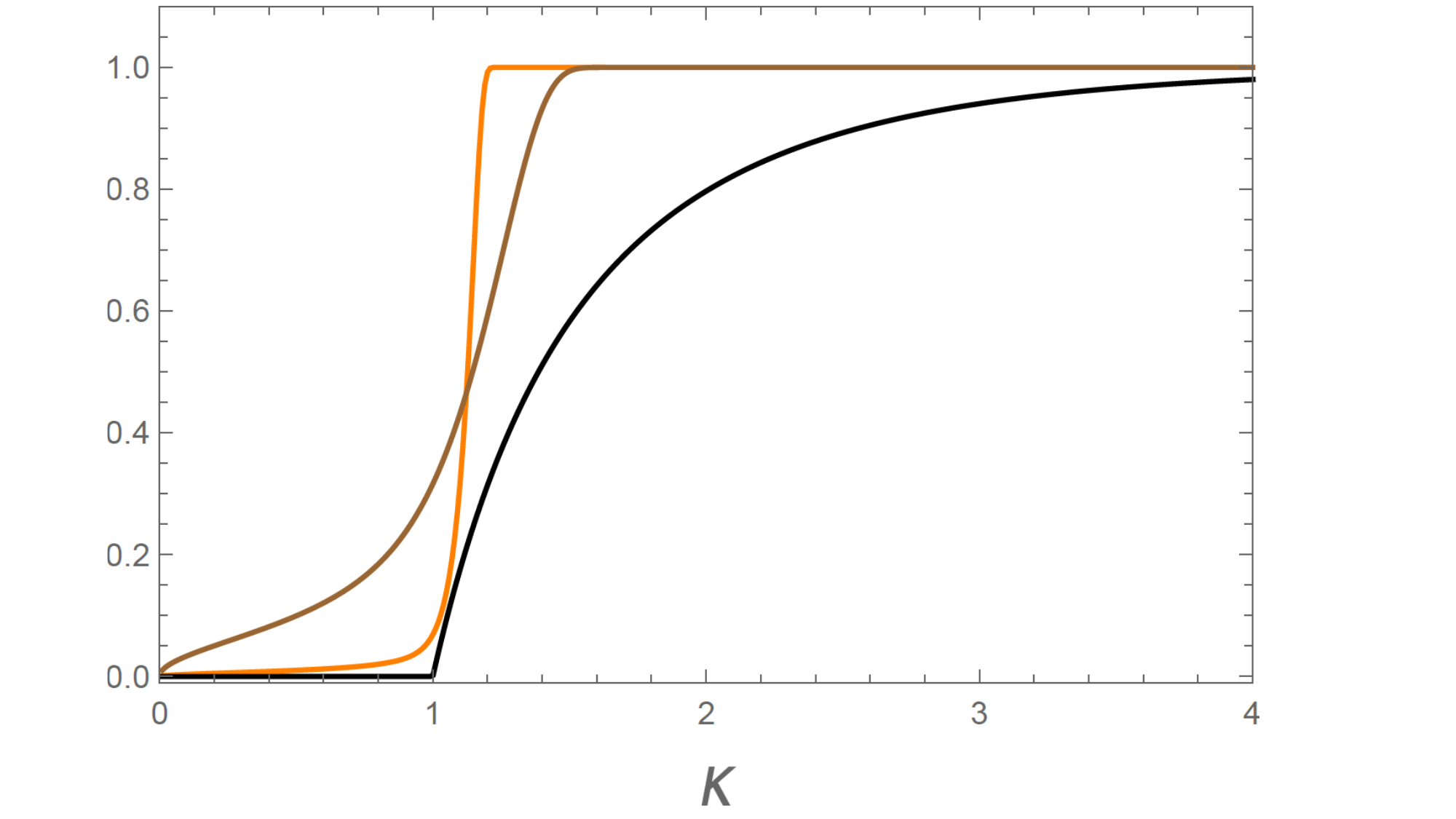}
		\caption{Comparison of $f_{\rm cluster}$ (black curve) with the commensurate mean field order parameter $(1 - P_\infty)^{1/2}$ for $N=10^2$ (brown) and $N=10^4$ (orange) \label{simk3}.}
	\end{center}
\end{figure}
As can be seen from Fig. \ref{simk3} we see that the mean field results appear to get the nature of the phase transition wrong by suggesting a first-order transition (given the evident discontinuity developing in the thermodynamic limit) rather than the second-order transition suggested by the analytic approximation for $f_{\rm cluster}$. This failure of the mean field approximation is neither surprising nor alarming, and merely highlights its expected failure whenever fluctuations invalidate the mean field treatment. Nevertheless, the approximation ought to improve rapidly as one moves away from the critical point, which we now investigate numerically. \rd{It should come as no surprise that when the network consists of a giant connected component, the mean field approximation works well given the high dimensionality of system set by $N$.}

\begin{figure}
	\begin{center}
		\includegraphics[height=3.0in, width = 5.8in]{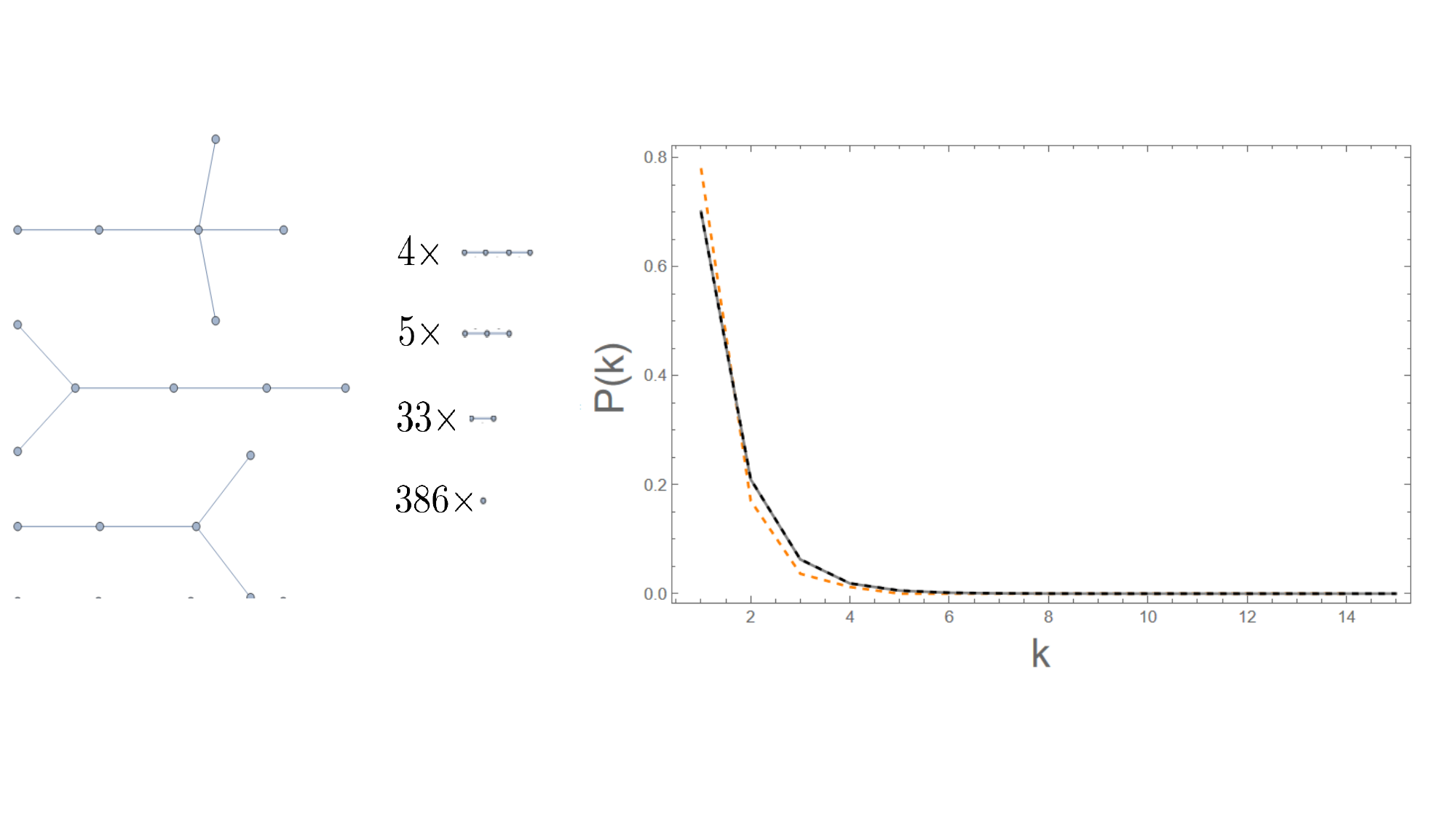}
		\caption{Comparison of simulated network (orange dashed) with the normalized mean field approximation (gray) and the exact answer given by Eq. \ref{scea} (black dashed, indiscernible from mean field result) for $p = 0.3/(N-1)$; $N = 500$. Graph on left is a representative draw from the ensemble.  \label{simM0p3}}	
	\end{center}
\end{figure}
\begin{figure}
	\begin{flushleft}
		\includegraphics[height=3.0in, width = 6.0in]{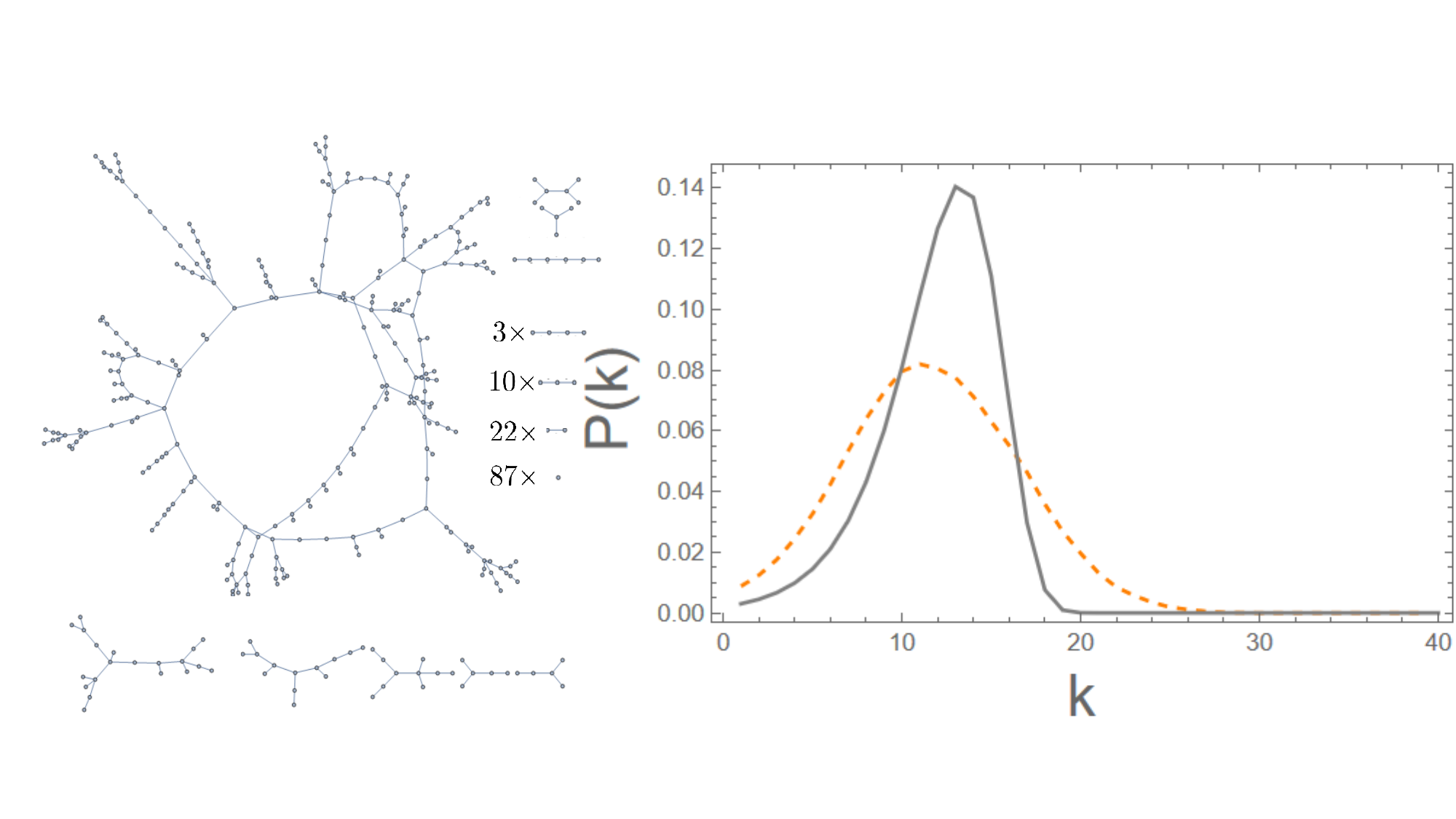}
	\caption{Comparison of simulated network (orange dashed) with the mean field approximation (gray) for $p = 1.5/(N-1)$; $N = 500$. Graph on left is a representative draw from the ensemble. \label{simM1p5}}	
	\end{flushleft}
\end{figure}
\begin{figure}
	\begin{flushleft}
		\includegraphics[height=3.0in, width = 6.0in]{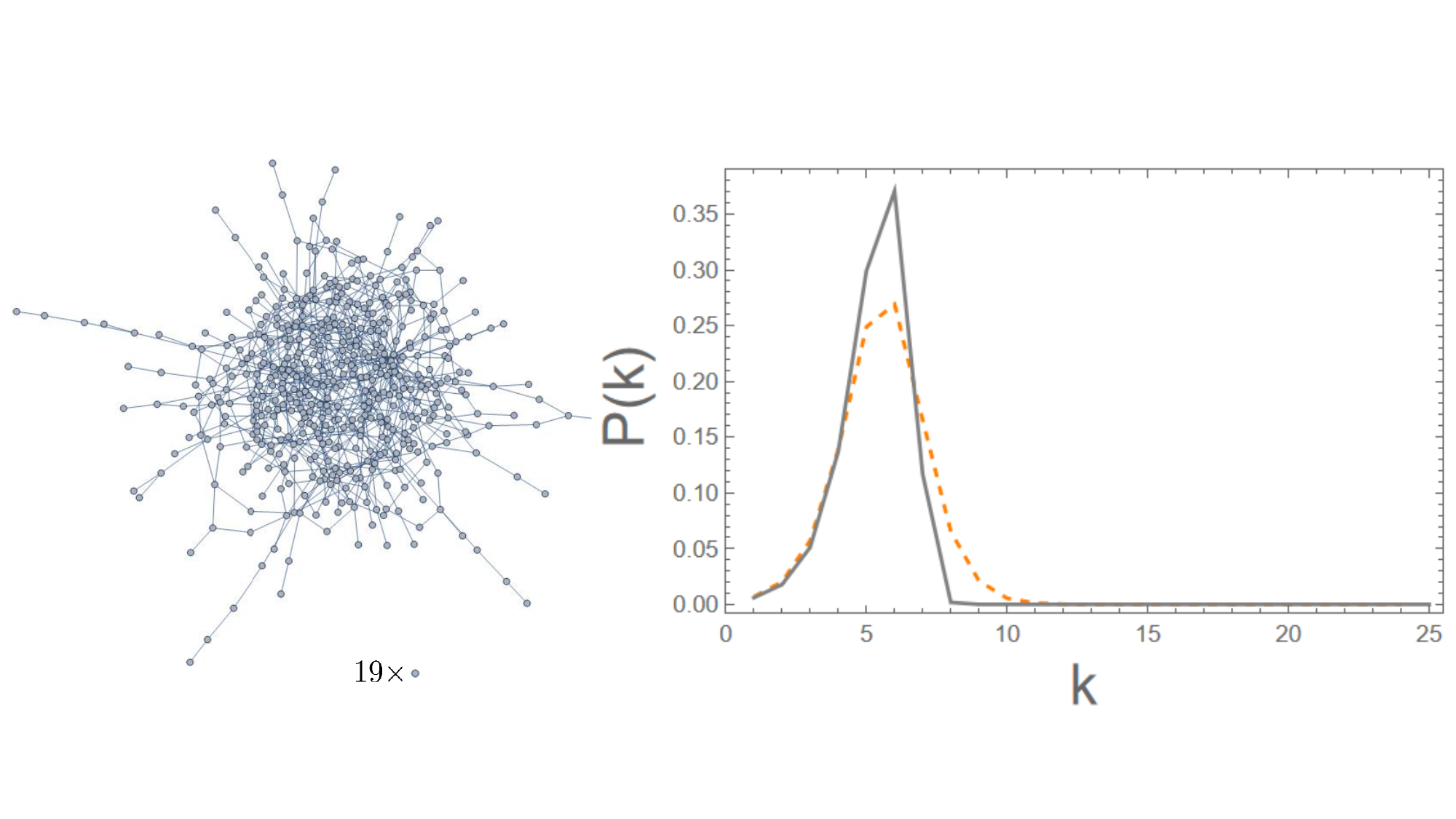}
		\caption{Comparison of simulated network (orange dashed) with the mean field approximation (gray) for $p = 3.0/(N-1)$; $N = 500$. Graph on left is a representative draw from the ensemble. \label{simM3}}	
	\end{flushleft}
\end{figure}


We performed a series of simulations of the PDF for the quantities $P(k)$ that describe the probability that the shortest path between two nodes will be of length $k$.  The results of such a check are shown in Figs. \ref{simM0p3} -- \ref{simM3} for an ensemble size of a 100 samples of a system of $N = 500$ nodes with $\kappa$ of 0.3, 1.5, and 3.0, respectively.  Here, we have chosen to neglect the single  term $P_{\infty}$.  Thus, the final term in Eq.\ \ref{norm} has been neglected and the remaining terms renormalized so that their sum is $1$.  For the cases $\kappa = 0.3$ and $3.0$, the system is  expected to be relatively homogeneous.  In the former case, $\kappa$ is simply too small for large clusters and thus large distances, $k$, to form. Although $P_{\infty}$ is not negligible, it has no effect other than the renormalization of the $P(k)$ shown in the figure.  In the latter case, we have seen that the largest cluster is formed predominantly by the merger of smaller clusters.  As can be seen from Eq. \ref{hermit}, all that remains is a small (i.e., $\exp[-\kappa]$) fraction of isolated nodes \rd{in the large $N$ limit}.  In both cases, the resultant homogeneity of the networks is relatively high, and the agreement between mean field and simulated results is reasonably good. \rd{In fact, for the sub-critical value of $\kappa = 0.3$, the exact distribution of shortest paths is known to be exponential, and given by \cite{katzav2018distribution}
\eq{scea}{P(k) = (1-\kappa)\kappa^{k-1},}
where we see from Fig. \ref{simM0p3} that the mean field is in excellent agreement with the former. That the simulation differs is an artifact of the finite size of the simulated system.} 

The case shown in Fig. \ref{simM1p5} is rather different since the value of $\kappa = 1.5$ has been chosen so that the largest cluster in the simulation contains half of the nodes.  The ensemble of adjacency matrices is expected to be maximally heterogeneous with considerable differences between its various elements.  In short, the mean field description should be at its worst.  Specifically, we should expect that the root-mean-square (rms) deviation of the $P(k)$ should be relatively large and that the agreement with mean field results should be rather poor. 

To conclude this section, we have presented a mean field description of the properties of an ensemble of randomly drawn adjacency matrices.  This description is sufficiently simple to be applied to networks of size $10^7$ or larger.  It correctly describes the qualitative transition of the system from an set of unconnected nodes to one dominated by a single small-worlds connected  network containing an increasing fraction of nodes as the number of links per node grows.  This result was obtained by making the conservative assumption that the average number of links per node, $\kappa$, is independent of the size of the system, $N$.   Any monotonic growth in $\kappa$ as a function of $N$ would necessarily result in a genuine small world involving the participation of all nodes in the thermodynamic limit $N \to \infty$.  Since many popular models build in small worlds behaviour by construction, we consider it appealing that random networks show similar behaviour generically.  As expected, the mean field description fails to capture the true nature of the small world phase transition.  Nevertheless, its description of the distribution of the distances between nodes appears to be quite reliable for all values of $\kappa$ considered here.  Thus, it is with some confidence that we can now turn to the more challenging and potentially more interesting topic of small worlds and dimorphic networks.

\section{Dimorphic Networks}
\begin{minipage}{0.5\textwidth}
	\begin{center}
		\includegraphics[height=2in, width = 3.5in]{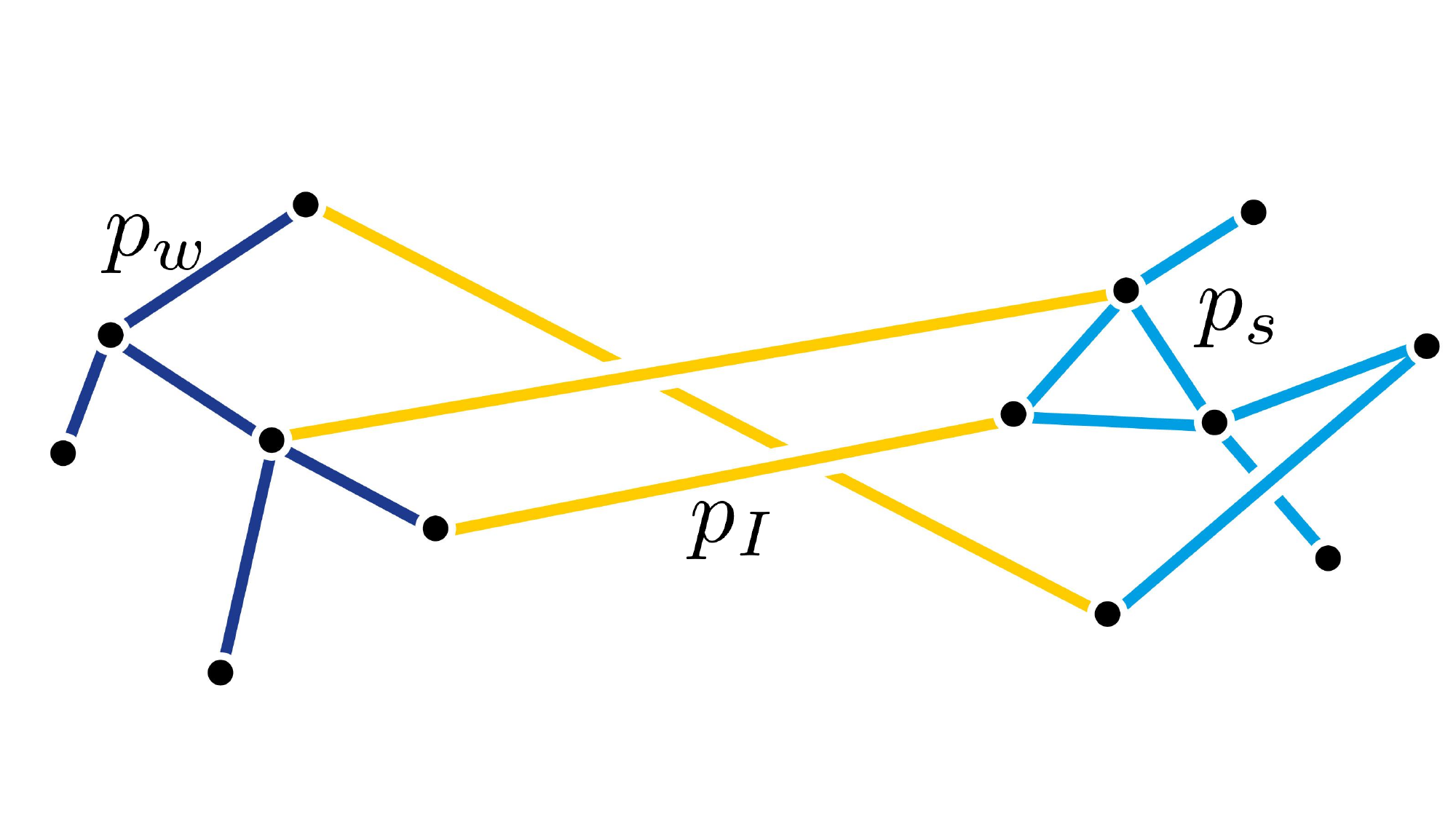}
	\end{center}
\end{minipage}
\begin{minipage}{0.5\textwidth}
	\begin{center}
		\eq{}{\nn~~\bf A = \left( \begin{array}{@{}c|c@{}}
				\begin{matrix}
					... & ... & ... & \\
					... & p_s & ... \\
					... & ... & ... 
				\end{matrix} 
				&\begin{matrix}
					... & ... & ... & \\
					... & p_I & ... \\
					... & ... & ... 
				\end{matrix}   \\
				\cmidrule[0.4pt]{1-2}
				\begin{matrix}
					... & ... & ... & \\
					... & p_I & ... \\
					... & ... & ... 
				\end{matrix} & \begin{matrix}
					... & ... & ... & \\
					... & p_w & ... \\
					... & ... & ... 
				\end{matrix}  \\
			\end{array} \right)}
	\end{center}
\end{minipage}

We now consider a two family stochastic block model in which $m$ out a total of $N$ nodes form links with each other with probability $p_s$ while the remaining $N-m$ nodes form links with each other with probability $p_w$. A \textit{planted partition} model would correspond to the situation where $p_s = p_w$, and the \textit{balanced planted partition} model would correspond to the situation where furthermore, $m = N/2$ \cite{amini2018semidefinite}. The probability that a link is formed between nodes in different families is $p_I$. That is, the network factorizes into two independent Erd\"os-Renyi-Gilbert sub-networks with associated probabilities $p_s$ and $p_w$ respectively, with bridge links between the sub-networks appearing with probability $p_I$. The subscripts $s$ and $w$ (strong and weak) are meant to suggest our eventual interest in the regime where one family is much more strongly connected than the other, although in what follows we allow them to be arbitrary\footnote{Another variant of the dimorphic network with interesting applications would be to allow the links within and between the different families to have different weights. The formalism developed here can readily be generalized to this case and will be the subject of a follow up investigation.}. 

It is convenient to partition the index $i \subset \{1,...,N\}$ into lower case Latin $a \subset \{1,...,m\}$ for nodes in the `strong' family and Greek $\alpha \subset \{m+1,...,N\}$ for nodes in the `weak' family, so that the corresponding random adjacency matrix is given by 
\begin{eqnarray}
p(A_{ab} =  1) &=& p_s\\
\nn p(A_{a\alpha} = 1) &=& p_I\\
\nn p(A_{\alpha\beta} = 1) &=& p_w\ .
\end{eqnarray} 
The generalization of the mean field approximation to the dimorphic network consists of calculating the probability that the shortest paths between two nodes $i, j$ will be of length $k$. For $k=2$ this can be done straightforwardly:
\begin{eqnarray}
p(A^2_{ab} = 0) &=& (1 - p_s^2)^{m-2}(1 - p_I^2)^{N-m}\\ \nn p(A^2_{a\beta} = 0) &=& (1 - p_sp_I)^{m-1}(1 - p_Ip_w)^{N-m - 1}\\ \nn p(A_{\alpha\beta} = 0)&=& (1 - p_I^2)^{m}(1 - p_w^2)^{N-m-2}\ .
\end{eqnarray}
In order to proceed for $k > 2$, it is useful to note that, for a string of adjacency matrix elements with the summation over the indices yet to be performed, the probability that this string vanishes is given by
\eq{ind}{ p(A_{ij_1}A_{j_1j_2}... A_{j_{k-2}j_{k-1}}A_{j_{k-1}j} = 0) = 1 - p_s^{\alpha_0}p_w^{\alpha_1}p_I^{k - \alpha_0-\alpha_1}~~~~{\rm( no~summation~over~indices)}}
where $\alpha_0$ counts the number of times a pair of indices are both draw from the set $\{a\}$ and $\alpha_1$ counts the number of times a pair of indices are both draw from the set $\{\alpha\}$. Therefore, the probability that there is a shortest path of length $k$ between any two nodes $i,j$ is given by
\eq{ans}{p_{k, {\{s, I, w\}}} = 1 - \prod_{\alpha_0 = 0}^{k}\prod_{\alpha_1 = 0}^{k - \alpha_0}\left(1 - p_s^{\alpha_0}p_w^{\alpha_1}p_I^{k - \alpha_0 - \alpha_1}\right)^{\Gamma^{k,\{s, I, w\}}_{\alpha_0\alpha_1}} }
where the subscripts $\{s, I, w\}$ denote whether the pair of nodes $i,j$ are both in set $a$, have mixed indices $a,\beta$ or are both in $\alpha$ respectively, and $\Gamma^{k,\{s, I, w\}}_{\alpha_0\alpha_1}$ counts the number of distinct ways in which the combination $\alpha_0, \alpha_1$ appears in the sum without any repeated indices. The challenge is to calculate the combinatorial factors $\Gamma^{k,\{s, I, w\}}_{\alpha_0\alpha_1}$ (see Appendix C), to determine the quantities $q_{k,s} := 1 - p_{k,s}$ etc. and to construct the analogues of Eq. \ref{Pk}   
\eq{}{P_{s}(k) := p_{k,s}\prod_{j=1}^{k-1} q_{j,s}~~~ 1 < k < N,}
with similar expressions for $P_{I}(k)$ and $P_{w}(k)$.  These correspond to the probability that there will be a shortest path of length $k$ beginning and ending in the strong linked sub-network ($P_{s}$), beginning and ending in the weak linked sub-network ($P_{w}$) or beginning in one and ending in the other ($P_{I}$).   (The construction of these three quantities --- the dimorphic analogues of the mean field approximation to the $P(k)$ given by Eq. \ref{Pk} --- is presented in Appendix B.) As we note in the appendix, the calculation of the requisite combinatorial factors above maps onto the problem of computing the numbers of consecutively aligned spins on a finite spin chain with fixed boundary conditions. As we show in the next section, dimorphic networks evidently exhibit a rich phase structure. This includes a phase where small worlds behavior is induced in the weakly connected sub-network even if it is not itself within the small-worlds regime.

\section{Phases of Small Worlds in Dimorphic Networks}

In the following, we consider the dimorphic network to consist of a `strong' sub-network with $p_s$ well within what would be the small-world regime if it were an independent network and vary $p_w$ and $p_I$.  We find three distinct phases for the network which can be classified as disordered small worlds, spinodal small worlds, and induced small worlds. For each of these, we compare the results of simulations with our mean-field approximation, which as we shall see, remains accurate in these distinct phases provided the network as a whole does not decompose into too many disconnected components. In what follows, we consider the natural parameterization for dimorphic network probabilities
\eq{}{p_s := \frac{\kappa_s}{m},~~ p_w := \frac{\kappa_w}{N-m},~~ p_I := \frac{\kappa_I}{N-m},}
where, as above, the full network consists of $N$ nodes, the strong sub-network contains $m$ nodes, and the weak sub-network contains $N-m$ nodes.

\subsection{Disordered small worlds}

\begin{figure}[!h]
	\includegraphics[height=4.0in, width = 6.5in]{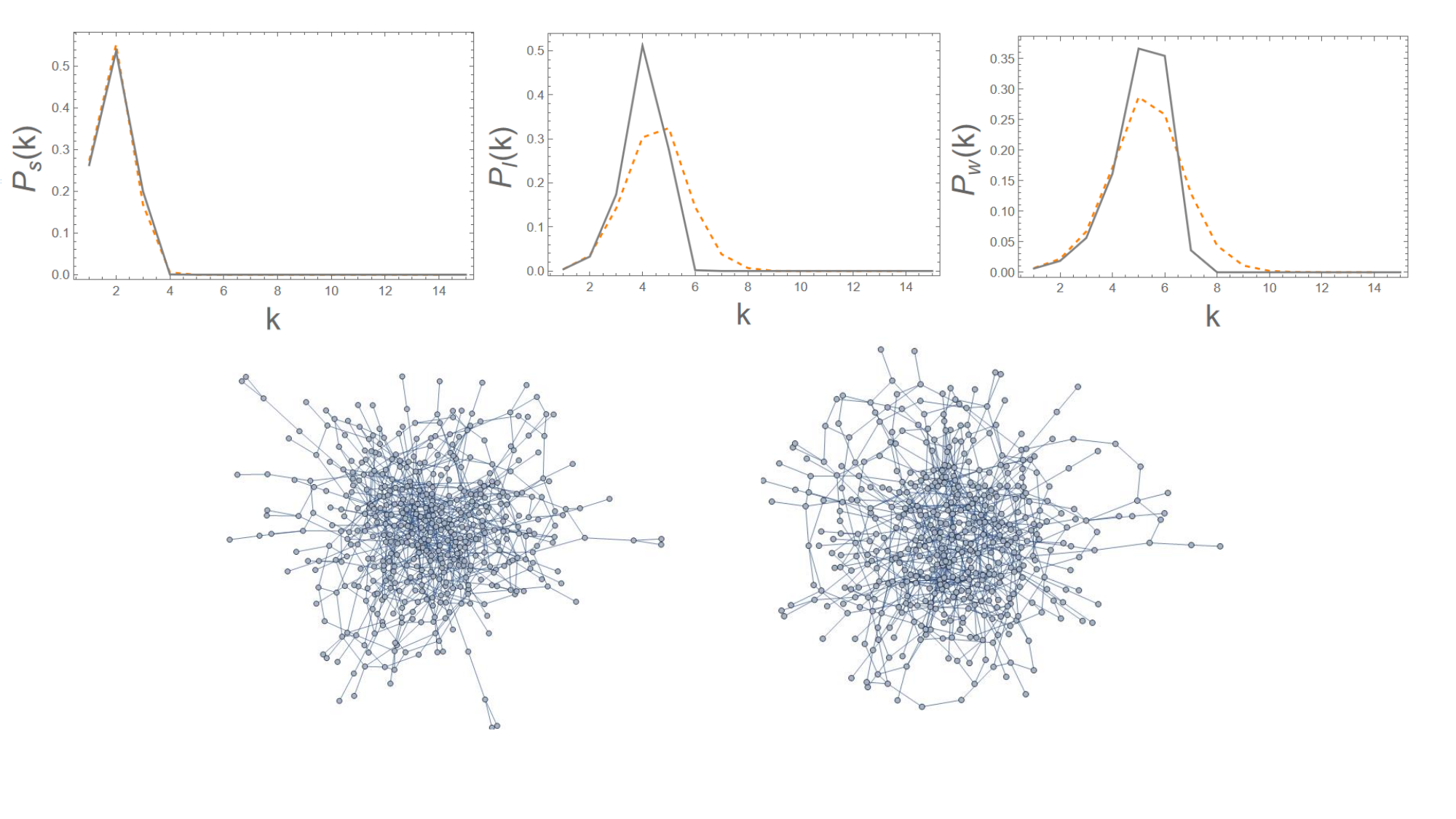}
	\caption{\label{disordered} Comparison of simulations (orange dashed) with the mean field approximation (gray):  $P_{s}(k), P_{I}(k)$ and $P_{w}(k)$ for $\kappa_s = 5$,  $\kappa_w = 2$ and $\kappa_I = 3$ for $m = 20, N = 500$. Depicted below are two representative largest connected component draws from the simulation ensemble. }
\end{figure}

In this regime, we consider $\kappa_s \neq \kappa_w \neq \kappa_I$ such that individually, each $\kappa_{i} \gtrsim 2$, chosen to keep the network fragmentation to a minimum to ensure the validity of the mean-field approximation. In Fig. \ref{disordered}, we see that the mean-field approximation accurately compares to simulations. By varying $N$, one also finds the small world scaling of paths that begin and end with a given sub-network, or beginning in one and ending in another. The distinction between the sub-networks is not particularly relevant in this regime, and one can consider this network globally as a disordered small world, as can readily verified by increasing $N$ and $m$. 

\subsection{Spinodal small worlds}

In this regime, we consider $\kappa_s, \kappa_w \gtrsim 2$, with  $\kappa_s, \kappa_w  \gg \kappa_I > \kappa_{I_{\rm\, crit}}$, where we have defined
\eq{}{\kappa_{I_{\rm\, crit}} = \frac{1}{m}.}
With these parameters, the network decomposes into two strongly connected sub-networks with a small number of bridge links between them. The critical value for $\kappa_I$ can be understood as the value much below which the strong and weak sub-networks fail to form any links between them. Precisely at the critical value, any given realization of the network will be as likely as not to form bridge links between the strong and weak sub-networks. Were one to begin in the disordered small-worlds regime and decrease $\kappa_I$ until it is much less than $\kappa_s$ and $\kappa_w$, but still greater than $\kappa_{I_{\rm\, crit}}$, the network would undergo a transition roughly analogous to spinodal decomposition in a fluid, wherein two phases co-exist and the transition occurs instantaneously due to the absence of any nucleation or latent energy difference between the two phases.  

\begin{figure}[h]
	\includegraphics[height=4.0in, width = 6.5in]{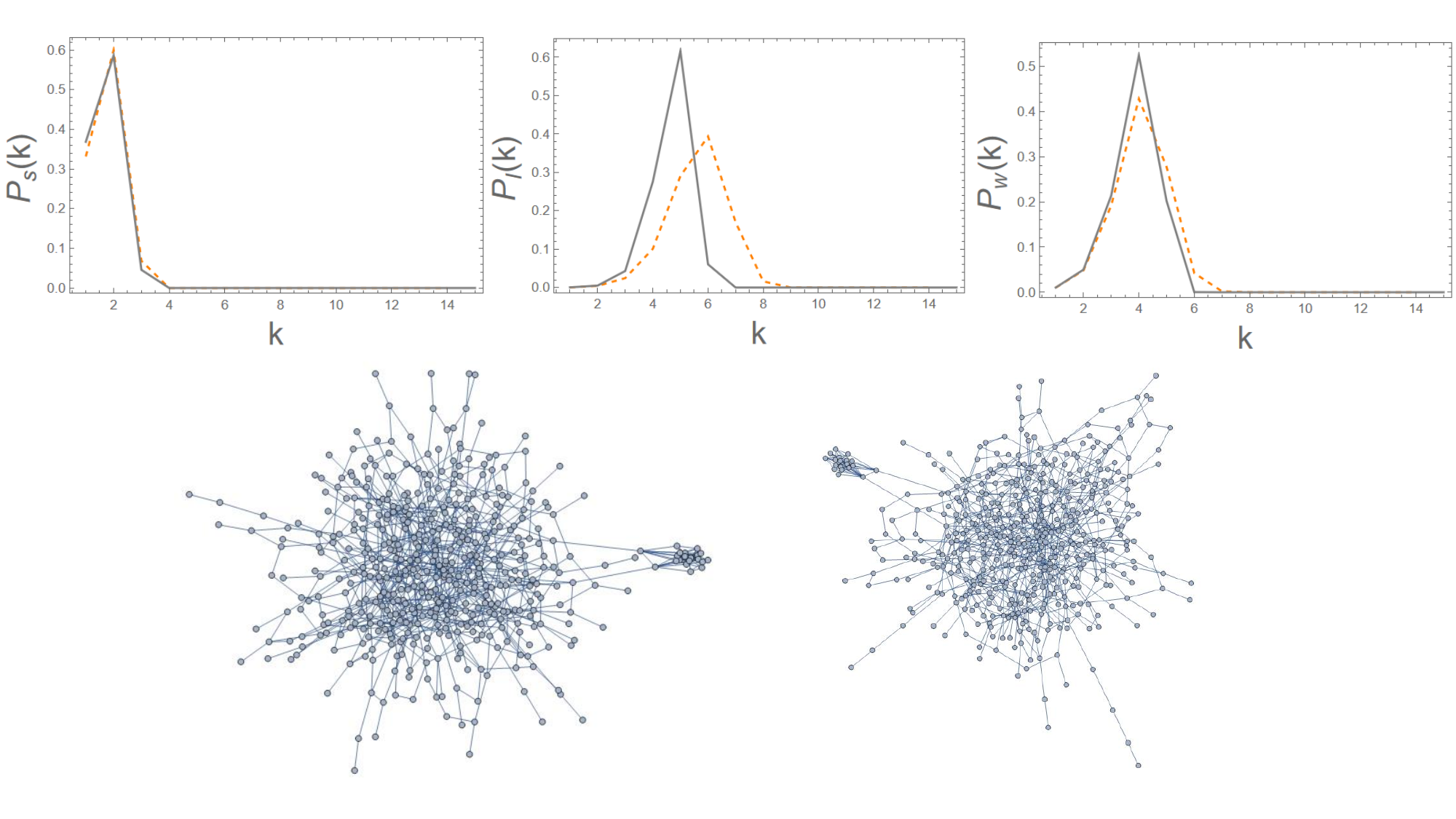}
	\caption{\label{spinodal} Comparison of simulations (orange dashed) with the mean field approximation (gray):  $P_{s}(k), P_{I}(k)$ and $P_{w}(k)$ for $\kappa_s = 7$,  $\kappa_w = 5$ and $\kappa_I = 0.2$ for $m = 20, N = 500$. Depicted below are two representative largest connected component draws from the simulation ensemble.}
\end{figure}

\begin{figure}[h!]
	\includegraphics[height=2.0in, width = 6.5in]{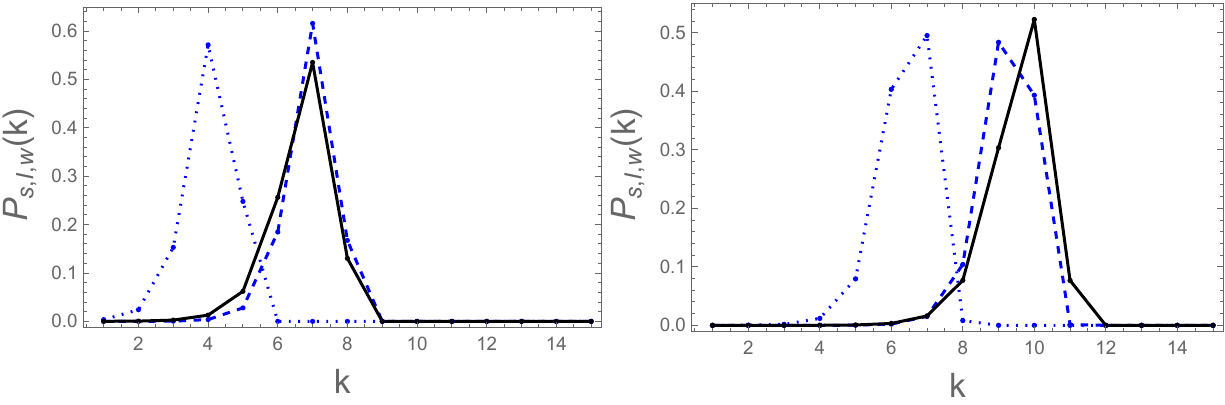}
	\caption{\label{spinodalplots} Scaling of $P_{s}(k)$ (blue dotted), $P_{I}(k)$ (blue dashed) and $P_{w}(k)$ (black) for $\kappa_s = 7$,  $\kappa_w = 5$ and $\kappa_I = 0.2$ for $m = 2\times10^3, N = 5\times10^4$ (left) and $m = 2\times10^5, N = 5\times10^6$ (right).}
\end{figure}

We again compare the results of our mean-field approximation to simulations in Fig. \ref{spinodal}, where we find good agreement for the mean and modes of the distributions. By virtue of the bridge links, one finds small world scaling in the spinodal regime for each type of path -- whether beginning or ending in a given sub-network, or beginning in one and ending in another, as depicted in Fig. \ref{spinodalplots}. Here, the ratio $m/N$ is kept fixed as $N$ is increased.  

\subsection{Induced small worlds} 
\begin{figure}[h]
	\includegraphics[height=4.0in, width = 6.5in]{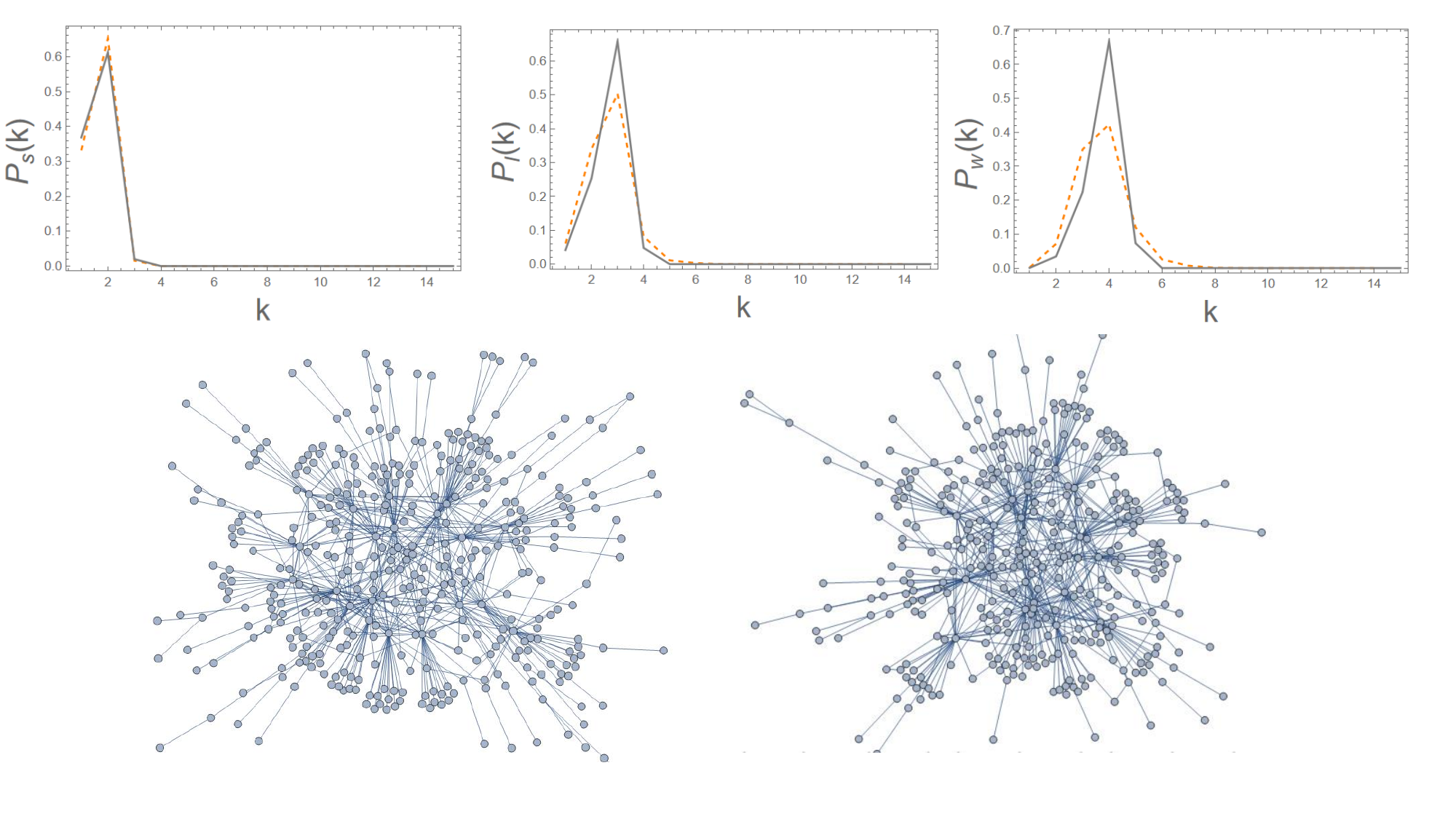}
	\caption{\label{induced} Comparison of simulations (orange dashed) with mean field approximation (gray):  $P_{s}(k), P_{I}(k)$ and $P_{w}(k)$ for $\kappa_s = 7$,  $\kappa_w = 0.3$ and $\kappa_I = 20$ for $m = 20, N = 500$. Depicted below are two representative largest connected component draws from the simulation ensemble.}
	\includegraphics[height=2.0in, width = 6.5in]{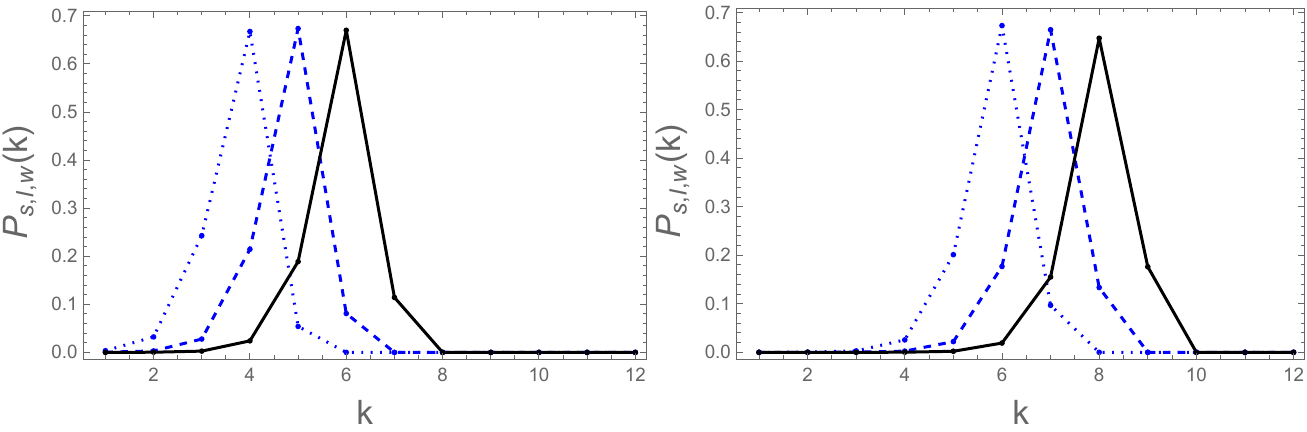}
	\caption{\label{inducedplots} Small world scaling of $P_{s}(k)$ (blue dotted), $P_{I}(k)$ (blue dashed) and $P_{w}(k)$ (black) for $\kappa_s = 7$,  $\kappa_w = 0.3$ and $\kappa_I = 20$ for $m = 2\times10^3, N = 5\times10^4$ (left) and $m = 2\times10^5, N = 5\times10^6$ (right). N.B. the difference of $+2$ between the mode of the distributions for $P_{w}(k)$ and $P_{s}(k)$.}
\end{figure}

In this regime, we consider $\kappa_s \gtrsim 2$, $\kappa_w < 1$ $\kappa_I \gg 1$ and once more compare the results of our mean-field approximation to simulations. In Fig. \ref{induced}, we see that even when the weak sub-network is not expected to be in the small world regime, the presence of bridge links with sufficient probability ($\kappa_I = 20$ in this example) connects the individual nodes of the weak sub-network to the strong sub-network such that small world scaling is recovered for nodes that begin and end in the weak sub-network. That is, small world behavior can be \textit{induced} in sub-networks that, by themselves, are not in the small world regime. This is straightforward to understand intuitively -- in the limit where the weak sub-network is completely disconnected, any bridge links will induce a connected sub-network out of the disconnected network. The average length for paths that begin and end in the connected part of the weak sub-network will thus be on average simply the average distance within the strong network + 2 for the hop back and forth. Although we require that the network not have too many disconnected components for the validity of the mean-field approximation, this simple reasoning does seem to hold for parameter values consistent with the validity of the mean-field treatment, as seen in Fig. \ref{inducedplots}.  

\section{Discussion}
\begin{figure}[h]
	\vspace{-20pt}
	\begin{flushleft}
		\includegraphics[height=3.0in, width = 6.0in]{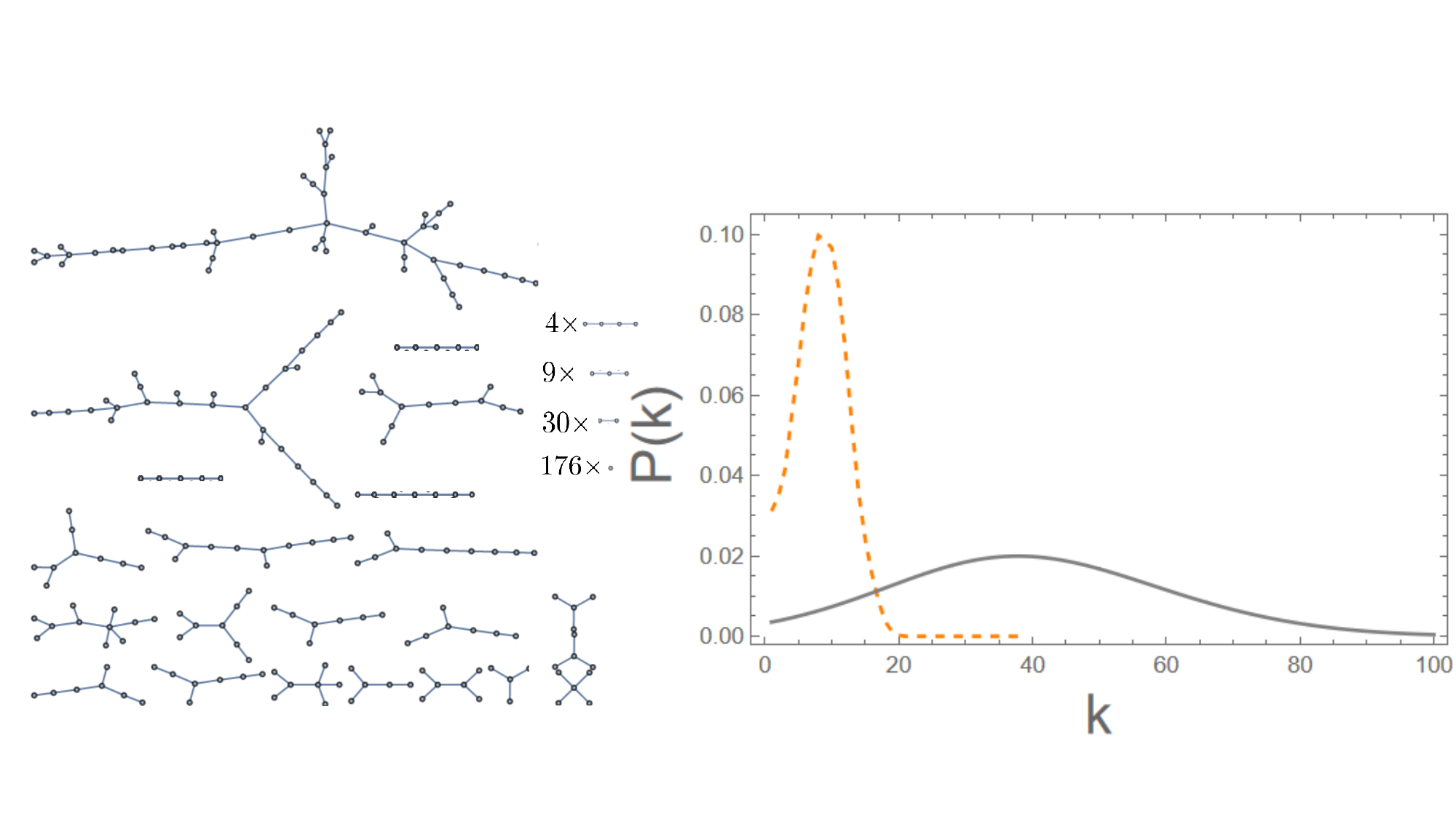}
		\caption{Approaching maximal failure of the mean field approximation close to the critical point at $\kappa = 1$: Comparison of simulated network (orange dashed) with mean field (gray) for $p = 1.1/(N-1)$; $N = 500$, along with representative draw from the graph ensemble. \label{simM4}}	
	\end{flushleft}
\end{figure}
In this paper, we introduced a novel mean-field approximation for random networks. The motivations for such an approximation is many-fold. A priori, one sees the immediate need for some sort of systematic approximation by first constructing the generating functional for arbitrary realizations of network randomness, from which ensemble averages of quantities of interest can be derived in principle. Even simple quantities of interest, such as the average of the elements of the distance matrix prove to be far too cumbersome to evaluate even in idealized examples. The motivation for the particular mean-field approximation introduced in section III was to explore the scaling behavior of average network distances, and that it should also be applicable to a wider variety of contexts. That is, even in cases where an underlying lattice sub-structure can't be appealed to, or when we introduce disorder in varying forms. 

The mean field approximation immediately yields a network order parameter $(1 - P_\infty)^{1/2}$ (Eq. \ref{Pinf}), requiring suitable interpretation. The existence of a critical point at $p = 1/(N-1)$ is manifest, however, as discussed in section III. One of the more predictable features of mean-field approximations its failure when ensemble fluctuations can no longer be neglected. As reasoned in section III, this is precisely the case as one approaches the phase transition, as depicted in Fig. \ref{simM4}\footnote{\rd{As remarked in Sect. III, the mode of our mean field approximation for the shortest path lengths diverges as $1/(\kappa - 1)$ as $\kappa \to 1$ from above, as found for the average path lengths in \cite{katzav2018distribution}.}}, and in Figs. \ref{simM0p3} - \ref{simM3}. 

The utility of the mean-field approximation is manifest in its generalizability to various forms of network disorder. The dimorphic variant of the stochastic block model studied in section IV exhibits a rich phase structure. It is intended as a model of network disorder where not all nodes are the same but can connect to other nodes with varying probabilities. The machinery introduced here suggests an immediate generalization to weighted networks where the dimporphism is extended to the weights and will be the subject of a followup study \cite{ADJSPP}. Furthermore, these results can in principle extend to stochastic block models with more than two families. As discussed in the appendix, this requires the calculation of combinatorial factors obtainable by constructing the generating function for auto-correlations over an $n-$symbol alphabet, where $n$ is the number of families. Our ultimate aim, is to arrive at as analytic understanding of various features of a prototype for disordered Gilbert-Erd\"os-Renyi networks and to apply the insights gained to various biological, sociologial and epidemiological systems.

\section{Acknowledgements}
We extend our warmest thanks to Diego Garlaschelli for his detailed comments on the manuscript and his many useful suggestions for its improvement. We'd also like to thank our anonymous referees for helping us substantially improve our discussion and referencing. 
\vspace{10pt}

\noindent\textit{Data availability statement: Data sharing not applicable to this article as no datasets were generated or analyzed during the current study.}

\appendix

\section{The Watts-Strogatz model}
The Newman-Watts variant of the Watts-Strogatz random network \cite{1998Natur.393..440W}\cite{NW} corresponds to a ring of nodes with links between every node and it's $m^{th}$ nearest neighbors, with random shortcuts inserted between any two non-adjacent nodes with probability $p$. For the case $m = 1$, the corresponding probability measure is given by 
\begin{eqnarray}
\label{NW}
(i + 2\leq j)~~~~~~ P_{ij} &=& p\,\delta(A_{ij} - 1) + (1-p)\,\delta(A_{ij})\\ \nn
(i, j = i+1)~~~~ P_{ij} &=&   \delta(A_{ij} -1 )\\ \nn
(i = j)~~~~~~~~~~~ P_{ij} &=& \delta(A_{ii})\\ \nn 
(i > j)~~~~~~~~~~~ P_{ij} &=& \delta(A_{ij} - A_{ji})~~~~~~~~ ({\rm Newman-Watts}~ m = 1)
\end{eqnarray}
where the indices are taken to be periodic. The above straightforwardly generalizes to Newman-Watts models with $m$ nearest neighbors. The generating function for Newman-Watts random graphs with $k=1$ follows from Eq. \ref{NW} and Eq. \ref{zans}:
\begin{eqnarray}
\label{NWans}
Z[J] &=& \prod_i e^{-J_{(i\,i+1)}}\prod_{i+1 < j}\left[p e^{- J_{(ij)}} + (1-p) \right]~~~~~~({\rm Newman-Watts}~ k = 1)\\ \nn
&=& e^{-{\rm Tr}J(V + V^T)}(1-p)^{\frac{(N-1)(N-2)}{2}}\prod_{i+1 < j}\left[1 + e^{-\left(\lambda + J_{(ij)}\right)} \right]
\end{eqnarray}
where $V$ is the $N$-dimensional permutation (or clock) matrix \cite{brualdi2006combinatorial}. From this, we find that
\eq{}{W[J] = \frac{(N-2)(N-1)}{2}\log\left(\frac{1}{1-p}\right) - \sum_{i + 1 < j} \log\left[1 + e^{-(\lambda + J_{(ij)})} \right] + {\rm Tr}J(V + V^T).}

\section{Mean field approximation for dimorphic networks}

The dimorphic network can be viewed as two separate component Gilbert networks with respective probabilities $p_s$ and $p_w$ for links within the component networks, with inter-component `bridge links' with probability $p_I$. The adjacency matrix thus decomposes into blocks, with the probability that any element within the diagonal blocks being unity given by $p_s$ and $p_w$ respectively, with probability $p_I$ for the off-diagonal blocks. We are thus interested in calculating
\eq{sum}{p\left(\sum_{j_1 \neq i, j}\sum_{j_2 \neq j_1, i, j}... \sum_{j_{k-1} \neq j_{k-2}... j_1, i, j}\hspace{-20pt} A_{i j_1} A_{j_1 j_2}... A_{j_{k-1} j}  = 0 \right),}
with different choices for $i, j$ yielding $p_{k,s}$, $p_{k,I}$ and $p_{k,w}$. Given the block structure, it helps to rewrite the matrix products within the summations as matrix products over blocks, i.e. we write
\eq{}{A_{ij} = A^{AB}_{i_A j_B}} 
where $A,B$ run over the binary indices $0,1$ with $i_0 \subset \{1,...,m\}$ and $i_1 \subset \{m+1,...,N\}$. Hence the summation in Eq. \ref{sum} can be expressed as e.g. for $p_{k,s}$ as
\eq{00sum}{ A^{0 A_{j_1}}_{a j_1} A^{A_{j_1}A_{j_2}}_{j_1 j_2}... A^{A_{j_{k-1}}0}_{j_{k-1} b} }
\textit{where the sum over the $A_{j_i}$ indices is unrestricted} whereas those of the corresponding lower indices do not have any of the $j_i$ repeated. For $p_{k,I}$ and $p_{k,w}$ the first and last (unsummed) indices in the above are assigned the values $0,1$ and $1,1$ respectively. For any given values of the lower indices, we have $p(A^{00}_{ab} = 1) = p_s$, $p(A^{01}_{a\alpha} = 1) = p_I$, and $p(A^{11}_{\alpha\beta}  = 1)  = p_w$. Clearly, as one sums over blocks in Eq. \ref{00sum}, the $A_{j_i}$ indices run over values that when viewed as a consecutive string, scan over all the integers between $0$ and $2^{k-1}$ represented as binary numbers. For $p_{k,s}$, we then affix $0$ to the beginning and end of each binary string, from which it should be clear from Eq. \ref{ind} that $\alpha_0$ in Eq. \ref{ans} corresponds to the number of consecutive 0's, and $\alpha_1$ corresponds to the number of consecutive 1's in any given resulting binary string. Let's take $k=3$ as an illustrative example -- the summation over the block indices for $p_{k,s}$ results in 
\begin{eqnarray}
\label{00}
A^{0 A_{j_1}}_{a j_1} A^{A_{j_1}A_{j_2}}_{j_1 j_2} A^{A_{j_{2}}0}_{j_{2} b} &=& A^{0\rdd{0}}_{a a_1} A^{\rdd{0} \bl{0}}_{a_1 a_2} A^{\bl{0} 0}_{a_2 b} \to (1 - p_s^3)^{(m-2)(m-3)}\\ \label{01}&+&
A^{0\rdd{0}}_{a a_1} A^{\rdd{0}\bl{1}}_{a_1 \alpha_2} A^{\bl{1}0}_{\alpha_2 b} \to (1 - p_sp_I^2)^{(m-2)(N-m)}\\ \label{10}&+&
A^{0\rdd{1}}_{a \alpha_1} A^{\rdd{1}\bl{0}}_{\alpha_1 a_2} A^{\bl{0}0}_{a_2 b} \to (1 - p_s p_I^2)^{(m-2)(N-m)}\\ \label{11}&+&
A^{0\rdd{1}}_{a \alpha_1} A^{\rdd{1}\bl{1}}_{\alpha_1 \alpha_2} A^{\bl{1}0}_{\alpha_2 b} \to (1- p_I^2p_w)^{(N-m)(N-m-1)}
\end{eqnarray}
where we color coded $A_{j_1}$ red and $A_{j_2}$ as blue. As we can see, the summation over the block indices when viewed as a string run over $00$, $01$, $10$ and $11$, i.e. from $0$ to $2^2$ in binary. Affixing 0's at the beginning and end of the string results in $0000$ (three consecutive pairs of 0's, so $\alpha_0 = 3, \alpha_1 = 0, k - \alpha_0 - \alpha_1 = 0$, cf. Eq. \ref{00}), 0010 and 0100 (one consecutive pairs of 0's, so $\alpha_0 = 1, \alpha_1 = 0, k - \alpha_0 - \alpha_1 = 2$, cf. Eqs. \ref{01} and \ref{10}) and 0110 (one consecutive pairs of 1's, so $\alpha_0 = 0, \alpha_1 = 1, k - \alpha_0 - \alpha_1 = 2$, cf. Eqs. \ref{11}). The remaining summation over distinct combinations of the lower indices yields the exponents in Eqs. \ref{00} - \ref{11}. Repeating the exercise affixing $0,1$ and $1,1$ to the beginning and end of the binary strings that run from $00 \to 11$ yields
\begin{eqnarray}
q_{3,s} &=& (1 - p_s^3)^{\sm{m-2}{2}}(1 - p_sp_I^2)^{2\sm{m-2}{1}\sm{N-m}{1}}(1 - p_s^2 p_w)^{\sm{N-m}{2}}\\
q_{3,I} &=& (1 - p_s^2p_I)^{\sm{m-1}{2}}(1 - p_sp_Ip_w)^{\sm{m-1}{1}\sm{N-m-1}{1}}\\ \nn &\times& (1 - p_I^3)^{\sm{m-1}{1}\sm{N-m-1}{1}}  (1 - p_I p_w^2)^{\sm{N-m-1}{2}}\\
q_{3,w} &=& (1 - p_I^2p_s)^{\sm{m}{2}}(1 - p_I^2p_w)^{2\sm{m}{1}\sm{N-m-2}{1}}(1 - p_w^3)^{\sm{N-m-2}{2}}
\end{eqnarray}
where $q_{k,s} := 1 - p_{k,s}$ etc., and where we've introduced the notation $\sm{a}{b} = a!/(a-b)!$. We can thus state the formal answer:
\begin{eqnarray}
q_{k, s} &=& \prod_{\alpha_0 = 0}^{k}\prod_{\alpha_1 = 0}^{k - \alpha_0}\left(1 - p_s^{\alpha_0}p_w^{\alpha_1}p_I^{k - \alpha_0 - \alpha_1}\right)^{\Gamma^{k,s}_{\alpha_0\alpha_1}\sm{m-2}{n^s_0}\sm{N-m}{n^s_1}}\\
q_{k, I} &=& \prod_{\alpha_0 = 0}^{k}\prod_{\alpha_1 = 0}^{k - \alpha_0}\left(1 - p_s^{\alpha_0}p_w^{\alpha_1}p_I^{k - \alpha_0 - \alpha_1}\right)^{\Gamma^{k,I}_{\alpha_0\alpha_1}\sm{m-1}{n^I_0}\sm{N-m-1}{n^I_1}}\\
q_{k, w} &=& \prod_{\alpha_0 = 0}^{k}\prod_{\alpha_1 = 0}^{k - \alpha_0}\left(1 - p_s^{\alpha_0}p_w^{\alpha_1}p_I^{k - \alpha_0 - \alpha_1}\right)^{\Gamma^{k,w}_{\alpha_0\alpha_1}\sm{m}{n^w_0}\sm{N-m-2}{n^w_1}}
\end{eqnarray}
where the combinatorial problem has been reduced to computing the different $\Gamma^{k,\{s,I,w\}}_{\alpha_0,\alpha_1}$, which count the number times $\alpha_0$ consecutive 0's and $\alpha_1$ consecutive 1's appear in the set of $2^{k-1}$ strings formed by writing $0 \to 2^{k-1}$ in binary (with each number being a string of $n_0$ 0's and $n_1$ 1's) and affixing 0's to the beginning and end of this representation for $p_{k,s}$, 0 and 1, and 1's for $p_{k,I}$ and $p_{k,w}$ respectively. Note that when we set $N=m$ and $p_s = p_I = p_w$ we recover the answer for a single component network found in the previous section. 

One can straightforwardly compute all the $\Gamma$'s via a recursion relationship derived in the next section of this appendix. Furthermore, the number of 0,1 indices summed over, given by $n_0, n_1$ are given by
\begin{eqnarray}
n^s_0 &=& \frac{1}{2}\left(k - 2 + \alpha_0 - \alpha_1\right)\\
n^I_0 &=& \frac{1}{2}\left(k - 1 + \alpha_0 - \alpha_1\right)\\
n^w_1 &=& \frac{1}{2}\left(k - 2 + \alpha_1 - \alpha_0\right)
\end{eqnarray}
where in all cases, we have $n_0^{\{s,I,w\}} + n_1^{\{s,I,w\}} = k-1$. As in Eq. \ref{Pk}, we are interested in computing
\eq{}{P_{s}(k) := p_{k,s}\prod_{j=1}^{k-1} q_{j,s}~~~ 1 < k < N,}
and similarly for $P_{I}(k)$ and $P_{w}(k)$. That is, $P_{s}(k)$ corresponds to the probability that there will be a shortest path of exactly length $k$ that begins and ends in the strong linked sub-network, $P_{w}(k)$ corresponds to the same but for a path that begins and ends in the weak linked sub-network, and $P_{I}(k)$ for paths that begin in one and ends in the other sub-network. 

\section{Multiplicity factors for dimorphic networks and spin chains}

The $\Gamma^{k,\{s,I,w\}}_{\alpha_0,\alpha_1}$  relate to an interesting question in the context of spin chains\footnote{What follows is a straightforward extension of the results contained in \cite{2016arXiv160900980D}.} -- it counts the number of ways one can have $\alpha_0$ pairs of consecutive spin pointing up and $\alpha_1$ pairs of consecutive spins pointing down on a chain of $k$ spins, with the boundary spins with specific orientations. In the present context, it relates to the quantities $z_s(n,l,m)$ defined as the number of ways in which $l$ pairs of consecutive zeroes and $m$ pairs of consecutive ones appear in a string of $n$ bits that begins and ends with $0$, and $z_I(n,l,m)$ defined as the number of ways in which $l$ pairs of consecutive zeroes and $m$ pairs of consecutive ones appear in a string of $n$ bits that begins with $1$ and ends with $0$. One could have defined a similar quantity for $z_w(n,l,m)$, where the string begins and ends with 1, but we notice that since flipping all the bits is an automorphism, it must be the case that $z_{w}(n,l,m) = z_s(n,m,l)$. Defining $z(n,l,m) = 0$ for $l+m \geq n$ and for all $n,l,m < 0$, we note that the following recursion relations hold:
\eq{zs}{z_s(n+1,l,m) = z_s(n,l-1,m) + z_s(n-1,l,m) + z_I(n-1,l,m-1)}
\eq{zI}{z_I(n+1,l,m) = z_s(n,l,m) + z_s(n-1,l,m-1) + z_I(n-1,l,m-2)}
For Eq. \ref{zs}, consider a string of length $n+1$ with $l$ pairs of consecutive 0's and $m$ pairs of consecutive 1's. The total number of these define the l.h.s. above. Consider now a given string that satisfies this criteria. Either it begins with 00 or 01. In the first case, it consists of a 0, followed by an $n$ string beginning and ending with 0, with $l-1$ pairs of consecutive 0's and m pairs of consecutive 1's (the number of these is the first term on the rhs of eq \ref{zs}). In the second case, it either begins with a 010 or a 011. If it begins with 010, it consists of 01, followed by an $n-1$ string beginning and ending with 0 with $l$ pairs of consecutive 0's and $m$ pairs of consecutive $1$'s, the number of which is the second term on the rhs of eq \ref{zs}. If it begins with a 011, it consists of 01, followed by an $n-1$ string beginning with 1 and ending with 0 with $l$ pairs of consecutive 0's and $m-1$ remaining pairs of consecutive $1$'s, the number of which is the third term on the rhs of eq \ref{zs}. Since this exhausts the possibilities, the equality follows. A similar argument results in eq \ref{zI}. It follows that 
\begin{eqnarray}
\Gamma^{k,s}_{\alpha_0,\alpha_1} &=& z_s(k+1,\alpha_0,\alpha_1)\\
\Gamma^{k,I}_{\alpha_0,\alpha_1} &=& z_I(k+1,\alpha_0,\alpha_1)\\
\Gamma^{k,w}_{\alpha_0,\alpha_1} &=& z_s(k+1,\alpha_1,\alpha_0)
\end{eqnarray}
One can generate all the appropriate degeneracy factors\footnote{In principle, there should also be a generating function for these symmetry factors by considering the problem of word auto-correlations over a two-symbol alphabet. Thus suggests that the generalization of the dimorphic network to a multimorphic setup, i.e. to a stochastic block model with more than two families can be obtained by determining these generating functions.} by straightforwardly iterating from the case for $k=2, 3$.

\bibliography{SW.bib}

\end{document}